\documentclass[twocolumn]{aastex701}

\usepackage{xspace}

\newcommand{\css}{\ensuremath{\mathrm{CS}_{2}}\xspace}
\newcommand{\coo}{\ensuremath{\mathrm{CO}_{2}}\xspace}
\newcommand{\soo}{\ensuremath{\mathrm{SO}_{2}}\xspace}
\newcommand{\methane}{\ensuremath{\mathrm{CH}_{4}}\xspace}
\newcommand{\water}{\ensuremath{\mathrm{H}_{2}\mathrm{O}}\xspace}
\newcommand{\ammonia}{\ensuremath{\mathrm{NH}_{3}}\xspace}
\newcommand{\sulfide}{\ensuremath{\mathrm{H}_{2}\mathrm{S}}\xspace}

\newcommand{\crat}{\ensuremath{^{12}\mathrm{C}/^{13}\mathrm{C}}\xspace}

\newcommand{\orat}{\ensuremath{^{16}\mathrm{O}/^{18}\mathrm{O}}\xspace}

\newcommand{\cmin}{\ensuremath{^{13}\mathrm{C}}\xspace}

\newcommand{\nmin}{\ensuremath{^{15}\mathrm{N}}\xspace}

\newcommand{\omin}{\ensuremath{^{18}\mathrm{O}}\xspace}

\newcommand{\hubble}{\textit{Hubble}}
\newcommand{\jwst}{\textit{JWST}}

\newcommand{\spitzer}{\textit{Spitzer}}
\newcommand{\tess}{\textit{TESS}}

\newcommand{\rearth}{\ensuremath{R_{\oplus}}}
\newcommand{\mearth}{\ensuremath{M_{\oplus}}}

\newcommand{\rjup}{\ensuremath{R_{\mathrm{J}}}}
\newcommand{\mjup}{\ensuremath{M_{\mathrm{J}}}}

\begin{document}

\title{Expanding the Volatile Abundance Inventory for High-Priority K-Dwarf JWST Exoplanet Systems}

\correspondingauthor{David R. Coria}

\author[orcid=0000-0002-1221-5346, gname=David, sname='Coria']{David R. Coria}
\affiliation{Department of Physics \& Astronomy, University of New Mexico, Albuquerque, NM 87131, USA}
\affiliation{Department of Physics \& Astronomy, University of Kansas, Lawrence, KS 66045, USA}
\email[show]{drcoria@unm.edu}

\author[orcid=0000-0001-5541-6087, sname='Hejazi']{Neda Hejazi} 
\affiliation{Instituto de Alta Investigaci{\'o}n, Universidad de Tarapac{\'a}, Santiago,  Chile}
\affiliation{Centro de Excelencia en Astrofísica y Tecnologías Afines, Santiago, Chile}
\email{nhejazi@gestion.uta.cl}

\author[gname=Ian,sname=Crossfield]{Ian J. M. Crossfield}
\affiliation{Department of Physics \& Astronomy, University of Kansas, Lawrence, KS 66045, USA}
\email{ianc@ku.edu}

\author[gname=Ismael, sname=Mireles]{Ismael Mireles}
\affiliation{Department of Physics \& Astronomy, University of New Mexico, Albuquerque, NM 87131, USA}
\email{mirelesi@unm.edu}

\author[gname=Casey, sname=Brinkman]{Casey L. Brinkman}
\affiliation{Trottier Space Institute, McGill University, 3550 Rue University, Montreal, QC H3A 2A7, Canada}
\email{casey.brinkman@mcgill.ca}

\author[gname=Jared, sname=Kolecki]{Jared R. Kolecki}
\affiliation{Department of Physics \& Astronomy, University of Notre Dame, Notre Dame, IN 46556, USA}
\email{jkolecki@nd.edu}

\author[orcid=0000-0001-8907-4775, gname=Emilio, sname=Marfil]{Emilio Marfil}
\affiliation{Departamento de Ingeniería Topográfica y Cartografía, E.T.S.I. en Topografía, Geodesia y Cartografía, Universidad Politécnica de Madrid, E-28031 Madrid, Spain}
\email{emilio.gomez.marfil@upm.es}

\author[gname=Thomas, sname=Nordlander]{Thomas Nordlander}
\affiliation{Theoretical Astrophysics, Department of Physics and Astronomy, Uppsala University, Box 516, SE-751 20, Uppsala, Sweden}
\email{thomas.nordlander@physics.uu.se}

\author[gname=Lauren, sname=Weiss]{Lauren M. Weiss}
\affiliation{Department of Physics \& Astronomy, University of Notre Dame, Notre Dame, IN 46556, USA}
\email{lweiss4@nd.edu}

\author[gname=Diana, sname=Dragomir]{Diana Dragomir}
\affiliation{Department of Physics \& Astronomy, University of New Mexico, Albuquerque, NM 87131, USA}
\email{dragomir@unm.edu}


\begin{abstract}

There are exceedingly few chemical abundance measurements available in the literature for K dwarf stars, especially for those with T$_{eff} <$ 4700 K due to an incompatibility of traditional analyses with their molecular-line-ridden optical spectra. To address the need for K dwarf abundance constraints, we present an elemental ([C/H], [O/H]) and isotopic (\crat, \orat) abundance analysis for six K dwarf exoplanet host stars: K2-141, NGTS-10, WASP-43, WASP-69, WASP-80, and HATS-72. Each system hosts an exoplanet targeted by JWST characterization efforts, making host star abundances especially critical as baseline references. Using MARCS model atmospheres, the TurboSpectrum spectral synthesis code, and near-infrared IGRINS spectra, we derive novel [C/H] and [O/H] abundance ratios for these high-priority K dwarf host stars. We further investigate prospects for determining \crat\ and \orat\ ratios using $^{13}$C$^{16}$O and $^{12}$C$^{18}$O lines from the 2.3 micron CO bandhead. The observed spectra are too noisy to measure \orat\ ratios for the sample, but we do find sub-solar \crat\ ratios for five out of six targets. We also derive uniform gyrochronological ages for each host star to compare with galactic chemical evolution models and identify underlying abundance-age trends. NIR H+K band spectroscopy provides an alternative to optical, atomic line-derived carbon and oxygen abundances for cool dwarf stars. The growing catalog of precise host star elemental and isotopic abundances complements ongoing observations of their companions and provides a baseline for understanding their chemical composition, origin, and evolution.

\end{abstract}



\section{Introduction}
Stellar chemical composition plays a central role in understanding planet formation, as stars and their planets form from the same protoplanetary disk. Higher bulk stellar metallicities are strongly linked to increased occurrence rates of close-in Jupiter-sized planets \citep{Valenti_Fischer_2005, Johnson_2010, Mortier_2013, Gan_2025}, and to a lesser extent, sub-Neptunes and super-Earths \citep{Wang_Fischer_2015, Petigura_2018, Boley_2024}. Weaker trends have also been reported for individual refractory elements, with subsolar-metallicity stars hosting gas planets showing enhanced refractory-to-iron abundance ratios compared with stars without detected planets \citep{Adibekyan_2012}.

The abundances of volatile elements such as C, O, N, and S in both stellar and planetary atmospheres provide insight into planet formation mechanisms and chemical evolution \citep{Bedell_2018, Brewer_Fischer_2016, Fortney_2012, Delgado_Mena_CO_2021, Turrini_2021, Crossfield_2023}. Carbon and oxygen abundances, in particular, can be used to differentiate between gravitational instability which yields stellar or super-stellar C/O ratios in brown dwarfs \citep{Ilee_2017, Hawkins_2020} while core accretion results in a more diverse range of C/O values in Jupiter-class exoplanets \citep{Xuan_2024b}. As an example, \citet{Hejazi2025} shows that the C/O ratio of the young M-type T Tauri star DH Tau A is fully consistent with that of its super-gas giant companion DH Tau Ab, suggesting that the companion likely formed through rapid gravitational collapse rather than a slower core accretion process. This formation scenario is further supported by the system's young age and relatively large companion-to-star mass ratio. 

The C/O ratio was once considered the ideal formation and migration tracer for exoplanets \citep{Oberg_2011, Madhusudhan:2012}: planets forming within the \water\ and CO snowlines retain stellar C/O ratios, while those forming farther out and migrating inward display a range of volatile enhancements for various reasons \citep{Madhusudhan_2014, Madhusudhan_2017, Schneider_Bitsch_2021a, Schneider_Bitsch_2021b, Bitsch_2022}. Today however, it is clear that a direct inversion from atmospheric abundances to planet formation models is quite difficult and such inferences rely heavily on our understanding of volatile delivery mechanisms and protoplanetary disk chemistry \citep{Mordasini_2016, Molliere_2022}. Planetesimal impacts and pebble accretion during planet formation greatly influence the final atmospheric composition of the gas giant exoplanets, as does the chemical evolution of the protoplanetary disk \citep{Mordasini_2016, Molliere_2022}. This demonstrates that chemical inheritance from the host star is only one area requiring careful consideration to connect planet formation mechanisms with observations of exoplanet atmospheres.

Recently, star-planet connection studies have been extended beyond the C/O ratio to include other volatile abundance ratios like C/N, N/O, and S/N which also offer diagnostics for a planet’s formation location and migration history \citep{Pacetti_2022, Oberg_2023}. In the directly imaged HR 8799 system, \cite{Xuan_2026} finds a tentative increase of planetary S/H with increasing orbital distance and also implicates \ammonia\ abundances as a formation location tracer relative to the CO and N$_2$ snowlines. Other studies suggest that N/S and C/O \citep{Nakazawa_2026} or C/S and O/S \citep{Crossfield_2023, Baburaj_2025} can disentangle various accretion scenarios (e.g. gas vs. solid, pebble vs. planetesimal accretion respectively) and are thus useful for testing planet formation models. This sort of testing requires complementary stellar abundance measurements to serve as a baseline for interpreting planetary chemical abundances \citep{Coria_2024, Baburaj_2025}. 

With powerful instruments like the James Webb Space Telescope and ground-based high-resolution spectrographs, astronomers can now detect even faint isotopologue lines from \cmin\ O, C\omin\, H$_2$\omin\, and \nmin\ H$_3$ in stars, brown dwarfs, and exoplanets \citep{Line_2021, Zhang_2021a, Zhang_2021b, Botelho_2020, Coria_2023, Gonzalez_Picos_2025, crossfield:2019a, Xuan_2024, Costes_2024, Hood_2024, de_Regt_2024, Gandhi_2023, Finnerty_2024, Gonzalez_Picos_2026}. These isotopic measurements, especially for \crat\ and \orat\ , provide an additional tool for distinguishing between formation pathways and tracing the chemical evolution of planetary systems \citep{Woods_2009b, Atreya_2016, Zhang_2021a, Zhang_2021b, Barrado_2023}.

\section{Motivation}
\subsection{K Dwarf Stars and Their Companion Exoplanets}
K dwarf host stars and their companion exoplanets are undeniably important to various aspects of exoplanet science, but they are not as widely studied as other dwarf stars. K dwarfs are more abundant than the best studied G-type stars, comprising $\sim12\%$ of the total stellar population compared with $\sim8\%$ for G stars \citep{Arney_2019}. One study finds that 4.0$\% \pm 2.3\%$ of late K dwarf stars (masses of 0.5-0.75 M$_{\odot}$ and effective temperatures of 3900-4800 K) host Saturn-mass or larger planets with orbital periods $<$ 245 days \citep{Gaidos_2013}. Furthermore, the sensitivity of exoplanet detection techniques like the radial velocity method and the transit method scales inversely with stellar mass and radius, such that smaller planets can be detected around smaller, less massive stars \citep{Gaidos_2013}. Radial velocity measurements rely on strong, numerous absorption lines that are prominent in the spectra of cool dwarf stars and scarce in hotter stars. The planet-to-star radius ratio ($R_{P}/R_{*}$) is also larger for planets orbiting smaller, cool dwarf stars which ultimately produces deeper, more detectable transits. K dwarf stars are also significantly less active than M dwarfs which not only makes these stars more amenable to RV and transit surveys \citep{Isaacson_Fischer_2010}, but also provides a more stable environment for companion planets \citep{Richey_Yowell_2019}. These positive attributes clearly make K dwarf host stars ideal candidates for characterization. So this raises the question, ``why are K dwarf stars so neglected"?

Late K-type stars in particular are left out from stellar, planetary, and chemical abundance surveys because they are faint and not M dwarfs. M dwarfs tend to spark more general interest because they are preferable targets for the search and characterization of habitable worlds due to their greater numbers and close-in habitable zones \citep{Gaidos_2013, Richey_Yowell_2019}. Most stellar surveys are flux- or magnitude-limited at visible wavelengths, meaning brighter, late F to early K spectral type stars are the optimal targets over late-type K and M dwarfs. Furthermore, deriving chemical abundances for cooler stars using traditional optical spectroscopy methods becomes more complicated as these stars contain millions of optical molecular lines, typically metal hydrides for K dwarfs and metal oxides for M dwarfs, that blend and obscure atomic signatures of interesting species. This is why established radial-velocity surveys like the California Planet Search (CPS) and the HARPS survey demonstrate a notable lack of late K stars \citep{Isaacson_Fischer_2010, Pepe_2000, Cosentino_2012} as do major optical chemical abundance surveys \citep{Brewer_2016, Bedell_2018}. Even state-of-the-art, optical EPRV spectrographs like ESPRESSO, EXPRES, and NEID observe only a handful of the hotter, early-type K dwarfs. Near-infrared instruments like CARMENES, SPIRou, and NIRPS, on the other hand, have a preference for M dwarfs instead with few dedicated K dwarf surveys like the KOBE (K-dwarfs Orbited By habitable Exoplanets) experiment \citep{Lillo-Box_2022, Balsalobre-Ruza_2025}. A similar challenge persists in IR spectra of cool dwarfs, where millions of background molecular lines, dominated by H$_2$O, blend with and obscure many potentially useful atomic and molecular abundance diagnostics. Even recent abundance surveys considering JWST host stars exclusively like \cite{Kolecki_2022} and \cite{Polanski_2022} covering Cycle 1 targets and \cite{Sun_2026} covering Cycle 2 targets do not include more than one target (18 Indi for \citet{Kolecki_2022}; WASP-69 for \citet{Polanski_2022, Sun_2026}) with effective temperatures $< 4800$ K in their sample. 

Stars hosting giant exoplanets are also considered high-priority targets because of the influence that giant exoplanets have on their planetary siblings. Gas giant exoplanets are thought to play a crucial role in the formation of their planetary systems as they facilitate the formation of planetary cores, alter protoplanetary disk chemistry by delivering volatile-rich planetesimals to the inner disk, and modify the influx of asteroids and comets to young planets \citep{Baygin_Laughlin_2015, Georgakarakos_2018, Bryan_2024}. Such investigations extend to multi-planet systems hosting warm-Jupiters as well. Warm Jupiters keep companions at rates comparable to smaller-planet systems, while hot Jupiters do not \citep{Huang_2016}. However, these companions almost always lie interior to the warm Jupiter’s orbit, and the habitable zone to date is unprobed at terrestrial-planet sensitivity.

Furthermore, gas giant exoplanets to date are more observable than their rocky exoplanet companions. Several ground-based observatories are dedicating significant time to the characterization of exoplanet atmospheres using high-resolution cross-correlation spectroscopy techniques \citep{Yakubu_2026}. Observing campaigns like the VLT/ESPRESSO GTO program, ESO/NIRPS GTO program, the ``Roasting Marshmallows" program with Gemini/IGRINS/IGRINS-2 \citep{Brogi_2023, Smith_2024_RM, Panwar_2025} and the ``ESO SupJup Survey" with VLT/CRIRES+ \citep{Gandhi_2025, Zhang_2024, Gonzalez_Picos_2025_GQLup, Gonzalez_Picos_2026}. reveal detections of various molecules, minor isotopologues, and other atmospheric properties. Likewise, \jwst\ observations have found a wealth of information on the composition, chemistry, structure, etc. of their atmospheres as discussed in the following section. Meanwhile, \jwst\ surveys of sub-Neptunes and smaller rocky exoplanets (e.g. the JWST COMPASS and Rocky Worlds programs) continue to reveal flat, featureless spectra-- in some cases expected for bare rock planets \citep{Adams_Redai_2025, Xue_2025, Lustig_Yaeger_2023, Kirk_2024, Scarsdale_2024, May_2023, Luque_2025} or otherwise flattened or muted by unocculted starspots, aerosol clouds/hazes or high mean molecular weight atmospheres \citep{Moran_2023, Meech_2026, Ahrer_2025, Alderson_2024, Wallack_2024, Alderson_2025, Teske_2025, Ahrer_2025, Alam_2025, Kempton_2023, Schlawin_2024_GJ1214, Ohno_2025, Roy_2026, Piaulet_Ghorayeb_2024, Mukherjee_2025}. In these cases, the nature of the atmospheres and interiors of terrestrial exoplanets are best inferred from the properties of their host stars and gas giant companions. 



\subsection{Elemental and Isotopic Abundances of K Dwarf Stars}
Despite significant progress over the past two decades in linking stellar chemical compositions to the properties of their orbiting planets, there remains a notable lack of such studies focused on low-mass stars—specifically late-type K and M dwarfs. These stars are intrinsically faint which means obtaining high-resolution, high signal-to-noise spectra suitable for abundance analyses requires substantial observing time on large, 8-meter-class telescopes. The challenges are compounded by the complex spectra of cool dwarf stars, which are dominated by molecular line blending, making precise abundance measurements especially difficult \citep{Hejazi_2023, Hejazi_2024}. As a result, only a limited number of studies report detailed elemental abundances for low-mass stars \citep{Souto_2017, Souto_2018, Souto_2022, Hejazi_2023, Hejazi_2024, Melo2024, Jahandar2025} in contrast to the extensive analyses and significantly larger samples available for hotter F, G, and early K-type stars \citep{Brewer_2016, Bedell_2018, Delgado_Mena_CO_2021, Kolecki_2022, Polanski_2022}.

Nevertheless, given the high occurrence rate of exoplanets around low-mass stars \citep{Gaidos_2013, Ment_Charbonneau_2023, Kaminski_2025}, expanding high-resolution spectroscopic studies of these hosts is essential for advancing our understanding of planet formation and composition. While recent efforts have begun to explore the chemical properties of hotter stars hosting high-priority JWST exoplanet targets \citep{Kolecki_2022, Polanski_2022}, a majority of high-priority host stars have little to no chemical abundance constraints in the literature. Nearly half of the JWST exoplanet host star sample consists of cooler stars with effective temperatures below 4700 K. Some of these targets have archival high-resolution spectra that could be useful for measuring chemical abundances, but many will have to rely on ongoing or future observations. As JWST and state-of-the-art ground-based observatories continue to deliver precise planetary abundance measurements, parallel spectroscopic analyses of cool dwarf host stars will provide critical context for understanding the processes that shape planetary atmospheres.

The paucity of elemental abundance constraints, even for high priority K dwarf exoplanet hosts, can be addressed moving forward using near-infrared H and K band spectroscopy covering wavelengths of 1.5 - 2.5 microns. There are significantly fewer chemical species accessible here; however, recent studies focused on deriving abundances for K and M dwarf stars have successfully constrained the abundances of up to 15 species (C, N, O, Na, Mg, Al, Si, K, Ca, Ti, V, Cr, Mn, Fe, Ni) for stars with effective temperatures $<$ 4700 K \citep{Souto_2022, Hejazi_2023, Hejazi_2024}. Notably, those abundance studies and earlier works like \citet{Tsuji_2015} utilize molecular OH, CO, and \water\ lines found throughout the H and K bands respectively to constrain carbon and oxygen abundance ratios crucial for contextualizing the yields from exoplanet atmosphere studies. 

Besides elemental abundances, near-infrared spectroscopy may also be the key to determining \crat\ and \orat\ ratios for cool dwarf stars. Recent studies have used lines in the fundamental (v = 1 to v = 0 transitions) rovibrational CO band located in the M band at 4.5 - 5.0 microns to derive \crat\ and \orat\ ratios for M dwarfs \citep{crossfield:2019a, Gonzalez_Picos_2026_TRAPPIST1} and solar twins \citep{Coria_2023} respectively. However, because K dwarfs are intrinsically faint, it is impractical to obtain sufficiently high signal-to-noise, high-resolution spectra at these wavelengths in a reasonable amount of telescope time, even on large, 8m-class telescopes. Another approach is to turn to the first overtone (v = 2 to v = 0 transitions) rovibrational band of CO located at 2.3 - 2.45 microns \citep{tsuji:2014, tsuji:2016a, tsuji:2016b, Gonzalez_Picos_2025}. Here the lines are weaker than in the M band, but K and M dwarfs are also brightest at these near-infrared wavelengths. Combined with the high-throughput of current near-infrared spectrographs like the Immersion GRating INfrared Spectrometer (IGRINS) \citep{Park_2014, Mace_2018}, it suddenly becomes more practical to obtain a sufficiently high signal-to-noise, high-resolution K band spectrum making this wavelength regime a viable option for measuring isotopic carbon and oxygen abundance ratios.

In this study, we use high-resolution ($R \sim 45,000$), high signal-to-noise IGRINS spectra to first constrain elemental carbon and oxygen abundances for the selected sample of K dwarf stars. Then, using these new elemental carbon and oxygen constraints, we investigate whether the isotopic abundance ratios, \crat\ and \orat, can also be inferred for these targets.

In Section \ref{sec:echoes_Kdwarf_sample}, we describe the K dwarf exoplanet hosts selected for this abundance analysis including an overview of their giant exoplanet companions and other relevant studies. In Section \ref{sec:echoes_obs_data}, we provide a brief description of the observations, spectral reduction, final stellar spectra used in this analysis. Then, in Sections \ref{sec:echoes_elemental_analysis} and \ref{sec:echoes_isotopic_analysis}, we detail the methodology used to generate synthetic spectra and measure elemental and isotopic abundance ratios for these K dwarf host stars. Finally, in Section \ref{sec:discussion}, we derive gyrochrone-based ages for our stellar sample and, along with the measured abundances, contextualize them with several galactic chemical evolution models.

\section{The K Dwarf Sample} \label{sec:echoes_Kdwarf_sample}
The cool dwarf stars selected for this elemental and isotopic abundance analysis are classified as K spectral type, main sequence stars with effective temperatures 4000 K $< T_{eff} <$ 4800 K. These stars also host exoplanets that are ideal candidates for atmospheric characterization via transmission spectroscopy, emission spectroscopy, or high-resolution cross-correlation spectroscopy \citep{Kreidberg_2018, Brogi_2019}, and have the most to gain from measurements of their host star's carbon and oxygen abundances. The majority are giant exoplanets with $R_P > 0.6 R_J$, but we also include a highly irradiated lava world to provide an alternative scenario for using host star abundances to contextualize exoplanet observations. Finally, the selected stars also have high-resolution, high signal-to-noise H+K band IGRINS spectra available through the Raw \& Reduced IGRINS Spectral Archive \citep{Sawczynec_2025} and via the following observing programs: GS-2023B-LP-108 (PI: Crossfield), GS-2024A-LP-108 (PI: Crossfield), and GS-2023A-Q-203 (PI: Crossfield).

The sample consists of six exoplanet host stars: K2-141, NGTS-10, WASP-43, WASP-69, WASP-80, and HATS-72. This includes exoplanets discovered by the Wide Angle Search for Planets (WASP) collaboration \citep{Pollacco_2006} that have been extensively studied for years but also some newer discoveries like K2-141 and NGTS-10 that are still not well characterized. The properties of the selected K dwarf host stars and their planetary companions are discussed further in the following section. Refer to Table \ref{tab:kdwarf_param_table} for a list of stellar parameters.


\begin{deluxetable*}{lcccccc}
\tablecaption{Stellar Parameters\label{tab:kdwarf_param_table}}
\tablewidth{0pt}
\tablehead{
\colhead{Parameter} & \colhead{K2-141} & \colhead{NGTS-10} & \colhead{WASP-43} & \colhead{WASP-69} & \colhead{WASP-80} & \colhead{HATS-72}
}
\startdata
Spectral Type & K7V & K5V & K7V & K5V & K7V & K4V \\
K Mag & 8.401 & 14.261 & 9.67 & 7.459 & 8.351 & 9.764 \\
G Mag & 10.7187 & 14.3145 & 11.8936 & 9.4915 & 11.2685 & 12.0646 \\
Parallax (mas) & 16.1303 & 3.8714 & 11.4740 & 19.8858 & 20.1141 & 7.8880 \\
Radius ($R_\odot$) & $0.68\pm0.02$ & $0.70\pm0.04$ & $0.65\pm0.01$ & $0.86\pm0.03$ & $0.59\pm0.02$ & $0.72\pm0.01$ \\
$^{*}T_{\mathrm{eff}}$ (K) & $4571\pm70$ & $4600\pm150$ & $4308\pm76$ & $4748\pm40$ & $4143\pm94$ & $4656\pm9$ \\
$^{*}\log g$ & $4.62\pm0.03$ & $4.59\pm0.02$ & $4.55\pm0.11$ & $4.50\pm0.07$ & $4.66\pm0.02$ & $4.59\pm0.01$ \\
$^{*}$[Fe/H] & $-0.03\pm0.08$ & $-0.02\pm0.12$ & $+0.12\pm0.12$ & $+0.24\pm0.02$ & $-0.13\pm0.17$ & $+0.10\pm0.01$ \\
$v\sin i$ (km/s) & $3.0\pm1.7$ & $4.0\pm0.6$ & $2.26\pm0.54$ & $2.20\pm0.40$ & $1.27\pm0.17$ & $0.80\pm1.30$ \\
References & (1, 7) & (2) & (3) & (4, 8, 9) & (5) & (6) \\
\enddata

\tablenotetext{}{Parameter search conducted using the NASA Exoplanet Archive \citep{Christiansen_2025}. Spectral types and $K$-band magnitudes from 2MASS \citep{cutri:2003}. $G$-band magnitudes and parallaxes from Gaia DR3 \citep{Gaia_2016, Gaia_2023}. (1)~\citet{Malavolta_2018}; (2)~\citet{McCormac_2020}; (3)~\citet{Esposito_2017}; (4)~\citet{Stassun_2017}; (5)~\citet{Triaud_2015}; (6)~\citet{Hartman_2020}; (7)~\citet{Barragan_2018}; (8)~\citet{Bonomo_2017}; (9)~\citet{Anderson_2014}. A star$^{*}$ indicates values adopted in this abundance analysis. Refer to Table~\ref{tab:stellar_abundances_Kdwarfs} for the original values and references.}

\end{deluxetable*}

\subsection{The K2-141 System}
K2-141 (EPIC 246393474) is a relatively bright (Kmag = 8.4), active K7V dwarf star about 68$\%$ the size of the Sun with an effective temperature of approximately 4571 K \citep{Malavolta_2018, Barragan_2018}. K2-141 is located $\sim 61$ pc away in direction of the constellation Pisces. This K dwarf star is host to K2-141 b, an ultra-short period (USP) super-Earth, and an outer companion K2-141 c, a Neptune-like planet, both discovered in 2018 by two independent studies using observations of the Kepler telescope during its extended “second light” mission, K2 \citep{Malavolta_2018, Barragan_2018}. 

K2-141 b orbits its host every 0.28 days (6.7 hours), while K2-141 c has an orbital period of 7.7 days \citep{Malavolta_2018, Barragan_2018}. Multiple transit and radial velocity studies studies have derived consistent parameters for this planet: a radius of approximately $R_p = 1.51 \pm 0.05$ \rearth\ and mass of $M_p = 5.08 \pm 0.41$ \mearth, leading to a high bulk density of around $8.2 \pm 1.1\,\mathrm{g\,cm^{-3}}$, indicative of a rocky, iron-silicate composition and the absence of a significant atmosphere \citep{Malavolta_2018, Barragan_2018}. K2-141 c's size is not well constrained due to its grazing transit and RV non-detections, but it is more likely a Neptune-like planet rather than a rocky super-Earth or hot Jupiter \citep{Malavolta_2018}.

Most Kepler and K2 ultra-short period planet candidates orbit stars too faint for precise radial velocity follow-up; however, K2-141's relative brightness allows for a precisely determined mass and thus a reliable density estimate \citep{Malavolta_2018}. K2-141 is also bright enough to conduct follow-up observations of the planet’s emission in the infrared as previously done with exoplanet 55 Cnc e, one of only a handful of other small USP planets accessible in both the optical and infrared \citep{Demory_2016a, Demory_2016b}. Although the larger planet K2-141 c is not a good candidate for characterization due to its grazing transits, K2-141 b's high surface temperature $>$ 2000 K makes it an excellent target for thermal emission observations as in the \spitzer\ analysis by \citet{Zieba_2022}.

The mineralogy of highly irradiated lava worlds is fundamentally shaped by both stellar carbon-to-oxygen and magnesium-to-silicon ratios, both in the planet’s deep interior and the observable atmosphere. Considering interior models of exoplanets, stellar C/O and Mg/Si ratios serve as the critical inputs used to break interior structure degeneracies, particularly silicate versus carbide-dominated interiors \citep{Dorn_2015, Dorn_2017}. The redox state and mineral composition of a rocky planet's mantle further control the chemistry of its outgassed atmosphere, meaning that the host star's C/O ratio leaves a fingerprint on the very gases a lava world releases under extreme irradiation \citep{Ortenzi_2020}. For example, magma ocean surfaces may produce chemically distinct vapor compositions depending on the rock type present \citep{Schaefer_Fegley_2009}, and such mantle outgassing drives measurable differences in atmospheric species like SiO, \coo, and \soo across different planetary compositions \citep{Kite_2016}. K2-141 b, as a highly irradiated rocky planet, is a prime testbed for mantle dynamics, nightside volcanism, volatile outgassing models, and the search for signatures of vaporizing surface rocks and minerals. Using two-dimensional convection models with tracer-based volatile tracking, studies like \citet{Meier_2026} find that magma ocean thickness varies from 200 to 300 km depending on the model parameters, corresponding to 2-3 $\%$ of the planet's radius. Nightside volcanism gradually depletes the mantle of volatiles resulting in the outgassing of tens of bars of CO$_2$ and H$_2$O over the course of a billion years. Notably, they demonstrate that even large volcanic eruptions on the nightside produce thermal emission signals of no more than 1 ppm which is well below the current detectability limits in thermal phase curves \citep{Meier_2026}. Nevertheless, the host star’s C/O ratio then provides crucial context for constraining the planet’s composition of the planet’s atmosphere and interior.

The JWST Cycle 1 proposal (GO $\#$2347, \citet{Dang_2021}) monitored K2-141 b for three complete orbits using MIRI/LRS, aiming to detect SiO in emission of this alleged lava planet, search for silicate clouds, and investigate the presence/structure of a global atmosphere. Analysis of the full-orbit spectroscopic MIRI/LRS (5-12 $\mu$ m) phase curve places constraints on the phase offset, Bond albedo, and heat recirculation efficiency from the phase curve, but ultimately inconclusive regarding the presence, or lack-of a global atmosphere on a lava planet \citep{Dang_2024}. The full results are not yet published.

\subsection{The NGTS-10 System}
NGTS-10 is a 10.4 $\pm$ 2.5 Gyr old K5V star with effective temperature T$_{eff} = 4400 \pm 100$ K and solar metallicity showing moderate signs of stellar activity. The star is named after its companion planet NGTS-10 b, the tenth discovery by the Next Generation Transit Survey (NGTS) at the ESO Paranal Observatory. The NGTS consists of an array of twelve 20 cm telescopes optimized for detecting small planets around K and early M stars. NGTS-10 b is an ultrashort period transiting hot Jupiter with a mass of 2.162 $^{+0.092}_{-0.107}$ \mjup, radius of 1.205 $^{+0.117}_{-0.083}$ \rjup, and orbital period of 0.767 days (18.4 hours) \citep{McCormac_2020}. As of its discovery date in 2020, NGTS-10 b is the shortest period hot Jupiter yet discovered, and joins the list of rare, ultrashort period (USP) hot Jupiters which includes notable, well-studied exoplanets such as WASP-18 b, WASP-19 b, WASP-103 b, HATS-18 b, KELT-16 b and WASP-43 b (also selected for this elemental and isotopic abundance study).

USP hot Jupiters are the easiest transiting planets to observe from the ground due to their relatively large transit depths and geometrically increased transit probability making them ideal targets for studying star–planet interactions and atmospheric characterization through transmission and emission spectroscopy \citep{McCormac_2020}. NGTS-10 b is likely to undergo strong tidal interactions with its host star due to its proximity at just 1.46 $\pm$ 0.18 Roche radii \footnote{the Roche radius is the distance from the host star where a self-gravitating planet disintegrates due to the tidal forces from the host}, which could trigger orbital decay and ultimately lead to its in-spiraling.

NGTS-10 b was observed by JWST in Cycle 1 through the GO Program Proposal 2158 \citep{Parmentier_2021}. This program measured a full 18-hour orbital phase curve of the hot Jupiter (0.6 - 5.3 $\mu$m) using JWST/NIRSpec. The study detects the presence of CO and H$_2$O with similar abundances on both dayside and nightside. Although the observations are consistent with a solar-composition atmosphere at chemical equilibrium for the dayside, the nightside favors disequilibrium chemistry due to its significant CH$_4$ depletion relative to equilibrium chemistry predictions \citep{Parmentier_2026}. The team argues that this is likely due to horizontal chemical quenching rather than non-solar elemental abundances or vertical mixing mechanisms. Most relevant to modeling the host star's spectrum, target acquisition imaging from this program revealed a stellar mass binary companion, NGTS-10 B, separated from the primary by ~ 0.33''. Although our target NGTS-10 A was initially modeled using the combined spectra from both components, the final effective temperature from this study ($T_{eff} = 4567 \pm 36$) is consistent with the previous estimated adopted here: $T_{eff} = 4400 \pm 100, 4600 \pm 150$ from Double SED fitting and HARPS spectroscopy, respectively \citep{McCormac_2020}. 

\subsection{The WASP-43 System}
WASP-43 is a K7V dwarf star with effective temperature T$_{eff} = 4500 \pm 100$ K and near-solar bulk metallicity and slightly super-solar metal abundances. The early discovery of hot Jupiter WASP-43 b, announced in 2011, makes WASP-43 one of the better characterized host stars in this sample. In fact, it is one of the targets of precursor spectroscopic studies for the ESO's Atmospheric Remote-sensing Infrared Exoplanet Large (Ariel) survey \citep{Magrini_2022, da_Silva_2024, Tsantaki_2025, Ramler_2026}. 

The companion planet WASP-43 b has been extensively studied by both ground-based and space-based observatories. WASP-43 b, as the name suggests, was discovered by the ground-based WASP collaboration \citep{Pollacco_2006}. The study found the planet transiting its host star every 0.81 days corresponding to a semi-major axis of only 0.014 AU or $2.1$ stellar Roche radii \citep{Hellier_2011}. The planet is about the size of Jupiter, but more massive with M = 1.8 \mjup, radius R = 0.9 \rjup, and a high bulk density \citep{Hellier_2011}. Like NGTS-10, WASP-43 is believed to undergo strong star-planet tidal interactions and possible decay of its orbit.

Hot Jupiters are among the easiest exoplanets to characterize from the ground due to their frequent transits and large transit depths, and this is certainly the case for WASP-43 b. Early transmission spectroscopy observations of WASP-43 sought to characterize its atmosphere both in the optical \citep{Murgas_2014} and near-infrared \citep{Chen_2014} but struggled to distinguish between a cloudy, high mean molecular weight, or hazy atmosphere due to large observational uncertainties. Another study focused instead on high-precision radial velocity measurements to measure the Rossiter-McLaughlin effect and refine orbital parameters like the spin-orbit angle \citep{Esposito_2017}. WASP-43 and WASP-80, both included in this elemental and isotopic abundance analysis, are among the coolest stars for which the Rossiter-McLaughlin effect has been observed so far. This study also found evidence for enhanced stellar activity possibly induced by the prominent star-planet interactions in the WASP-43 system. More recent ground-based studies focus on high-resolution, near-infrared cross-correlation spectroscopy of WASP-43 b. The hot Jupiter has been observed by both VLT/CRIRES+ and Gemini/IGRINS, the premier high-resolution, near-infrared spectrographs used for these types of analyses. The CRIRES+ retrieval reported the first detection of CO, provided evidence for the existence of \water, and measured a super-solar C/O ratio of 0.78 in the atmosphere of WASP-43 b \citep{Lesjak_2023}. Similarly, the IGRINS study detected \water\ but did not detect the presence of \methane, CO, or \coo\ in WASP-43 b's atmosphere \citep{Bartelt_2025}. Measurements of the C/O ratio of hot Jupiter atmospheres and their host stars are key to understanding both planet formation and migration but also cloud formation and atmospheric composition \citep{Reggiani_2022, Lesjak_2023}. 

Space-based observations of WASP-43 b have revealed differences between the planet's nightside and dayside. Thermal emission and transmission spectroscopy measurements obtained with instruments like \hubble\ allow for a precise determination of the water abundance in the atmospheres of hot Jupiters \citep{Kreidberg_2014}. Photometric reflected light observations of WASP-43 b using with \hubble\ reveal a very dark dayside atmosphere \citep{Fraine_2021}. Later studies using TESS and CHEOPS light curves of the planet argue this low dayside albedo is either due to a cloud-free atmosphere \citep{Scandariato_2022} or the presence of clouds containing strongly absorbing Fe-bearing aerosol species \citep{Chubb_2024}. Full-orbit phase curves from \spitzer\  are used to determine that WASP-43 b likely has a cloudy nightside that transitions to a relatively cloud-free dayside \citep{Murphy_2023}. A recent study observes WASP-43 b with JWST's MIRI instrument and finds a large day–night temperature contrast, evidence for water absorption, the presence of nightside clouds, and an absence of dayside clouds \citep{Bell_2024}. 

Because of WASP-43 b's numerous atmospheric abundance constraints, this planet is one of only a handful where a direct comparison can be made with its host stars' chemical abundances. This is essential to tracing a planet's formation, migration, and accretion histories and also to better understanding an exoplanet's atmospheric evolution.

\subsection{The WASP-69 System}
WASP-69 is an active, mid-K dwarf with effective temperature T$_{eff} = 4748 \pm 40$ K and super-solar bulk metallicity. WASP-69 is thought to be relatively young at an age of 0.8 - 1.1 Gyr estimated from its rotation rate, lithium abundance, and activity indicators \citep{Anderson_2014}. Like WASP-43, WASP-69 is another well-characterized system due to the early detection of a hot Jupiter companion orbiting the host. WASP-69 b is an inflated Saturn-mass planet (M = 0.26 \mjup, R = 1.06 \rjup) on a 3.868 day orbit \citep{Anderson_2014}.

Due to its proximity to its active host star and extended atmosphere, WASP-69 b is a favorable target for studies of atmospheric loss. Several high-resolution, near-infrared studies have targeted WASP-69 b in search of the metastable helium triplet line located at 1083.3 nm which is one of the best tracers of atmospheric escape in transiting exoplanets. This atmospheric escape is largely driven by the host star's x-ray and ultraviolet irradiation \citep{Sanz_Forcada_2025}. Studies using data from the ESO/NIRPS, Calar Alto/CARMENES, CFHT/SPIRou, and Keck/NIRSPEC spectrographs have confirmed excess absorption in the helium triplet during the transit of WASP-69 b and provided constraints on the planet's atmospheric mass-loss rate \citep{Nortmann_2018, Lampon_2023, Masson_2024, Allart_2025}. A recent Keck/NIRSpec study of WASP-69 b's atmospheric escape further finds that the atmospheric outflow extends at least 7.5 times the radius of the planet and corresponds to a mass-loss rate of 1 \mearth\ per Gyr \citep{Tyler_2024}.

Other ground-based studies of WASP-69 b find evidence for the presence of several key molecular species in the hot Jupiter's atmosphere. Near-infrared transmission spectroscopy reveals the presence of \methane, \ammonia, CO, C$_2$H$_2$, and \water\ indicative of a high C/O ratio \citep{Guilluy_2022} while optical transmission spectroscopy suggests the presence of TiO and high-altitude aerosols \citep{Ouyang_2023}. Even low-resolution spectrophotometry using VLT/FORS2 demonstrates evidence for \water and \methane absorption and high-altitude cloud decks while accounting for stellar faculae \citep{Petit_dit_de_la_Roche_2024}.

Recent observations of two WASP-69 b eclipses using JWST's NIRCam and MIRI instruments provide an emission spectrum covering 2 - 12 microns \citep{Schlawin_2024}. The data reveals absorption features from water vapor, carbon dioxide, and carbon monoxide, but no strong evidence for methane. Furthermore, WASP-69 b's emission spectrum seems inconsistent with cloud-free homogeneous models, instead favoring models with aerosols. The authors suggest possible model scenarios for the planet with the most plausible having higher metallicity and super-solar C/O: a cloud-layer model that includes high-altitude silicate aerosols to moderate long-wavelength emission; or a two-region model that includes significant dayside inhomogeneity and cloud opacity with two different temperature–pressure profiles \citep{Schlawin_2024}. New modules like \textit{NEXOTRANS} leverage Bayesian and machine learning techniques for retrievals while also incorporating disequilibrium chemistry, which when applied to the same JWST/NIRCam + JWST/MIRI spectra, place robust constraints on the volume mixing ratios of \water, \coo, CO, \methane, and potentially \soo\ \citep{Deka_2026}. 

Like WASP-43, WASP-69 b's numerous atmospheric abundance constraints make this planet one of only a handful where a direct comparison can be made with its host stars' chemical abundances to reveal the planet's formation, migration, accretion, and evolution.

\subsection{The WASP-80 System}
WASP-80, otherwise known as TYC 5165-481-1, is a relatively bright (Kmag = 8.4) late-type K dwarf star with sub-solar metallicity and effective temperature T$_{eff} = 4143 \pm 94$ K. The hot Jupiter WASP-80 b was discovered transiting this star in 2013 and has mass M $= 0.55 \pm 0.04$ \mjup, radius R $= 0.95 \pm 0.03$ \rjup\ and a circular, 3.07 day orbit \citep{Triaud_2013}. Like the other planets whose host stars were selected for this elemental and isotopic abundance study, WASP-80 b is another worthwhile target for characterization via transmission and emission spectroscopy.

From the ground, WASP-80 b has been observed in transmission by various high- and low- resolution spectrographs. At low-resolution, WASP-80 b's transmission spectrum is most consistent with a hazy atmosphere \citep{Kirk_2018}.  At high-resolution, WASP-80 b's transmission spectra reveal a mostly flat spectrum that favors high-altitude cloud models over cloud-free models with stellar or sub-stellar metallicities \citep{Parviainen_2018}. High-resolution, near-infrared transmission spectroscopy suggests an atmosphere consistent with solar composition and the presence of disequilibrium chemistry \citep{Carleo_2022}. This study also detects the presence of several molecular species in WASP-80 b's atmosphere including \water, \methane, \ammonia, and HCN with high significance, tentative detection of \coo, and inconclusive results for C$_2$H$_2$ and CO \citep{Carleo_2022}. Using the same data, another study looks for planetary absorption at He I triplet tracer of atmospheric escape but does not find any signature \citep{Fossati_2022}.

From space, WASP-80 b has been studied both by \spitzer\ and \jwst. Analysis of early \spitzer\ phase curves again show a flat transmission spectrum but do reveal WASP-80 b is as bright and as red as a T4-class brown dwarf, but that its temperature is much cooler \citep{Triaud_2015}. JWST/NIRCam transmission and emission spectra spanning 2.4-4.0 microns show strong evidence for \methane\ which are consistent with solar to sub-solar C/O and around five times solar metallicity for WASP-80 b's atmosphere \citep{Bell_2023}. A comparison of the shorter wavelength eclipse spectrum obtained with JWST's NIRISS/SOSS instrument with aerosol models demonstrates that the spectrum favors cloud species with weak-to-moderate near-infrared reflectance, soots, and hazes over manganese sulfide and silicate clouds \citep{Morel_2025}. More recently, a 2.4 - 10 $\mu$m transmission spectrum obtained by JWST/NIRCam and MIRI reveals evidence for \water, \methane, \coo, \ammonia, and CS$_2$ in the atmosphere of WASP-80 b and place upper limits on CO and SO$_2$ \citep{Triantafillides_2026}. This study establishes CS$_2$ as an observable tracer of sulfur disequilibrium chemistry that is also consistent with recent chemically validated networks that include efficient carbon-sulfur coupling through CH$_2$S.

WASP-80 b is yet another of only a handful where a direct comparison can be made with its host stars' chemical abundances. In this case, however, the limiting factor is the lack of chemical abundances reported for its cool, late-K-type host star WASP-80. 

\subsection{The HATS-72 System}
HATS-72 is a main sequence K dwarf with super-solar metallicity and effective temperature T$_{eff} = 4656 \pm 9$ K. HATS-72 b was discovered alongside three other giant planets transiting K dwarf stars-- HATS-47 b, HATS-48 b, and HATS-49 b-- by the HAT-South (HATS) collaboration \citep{Bakos_2013, Hartman_2020}. This hot Jupiter, also known as  WASP-191 b or TOI 294.01, was detected independently by the WASP collaboration and NASA's Transiting Exoplanet Survey Satellite  (\tess; \citealp{Ricker_2015}). HATS-72 b has a mass M $= 0.1254 \pm 0.0039$ \mjup, R $= 0.7224 \pm 0.0032$ \rjup, and an orbital period of 7.3279 days \citep{Hartman_2020}. Unlike the companion planets to the other host stars selected for this abundance study, HATS-72 b is classified as a warm-Jupiter rather than a hot Jupiter due to its longer orbital period $P = 7.33$ days. The expected equilibrium temperature of the companion planet is below 1000 K because of its host star's relatively small size and luminosity \citep{Hartman_2020}. 

HATS-72 b was observed by JWST's NIRSpec and NIRISS/SOSS instruments as part of the \jwst\ Cycle 1 GO Program 2594 \citep{Spake_2021}. The proposal aimed to test the``\methane thermometer" identified by \citet{Fortney_2020} which establishes that atmospheric \methane/\water and \methane/\coo abundance ratios are good probes of the interior temperatures of moderately-irradiated exoplanets (T$_{eq} \sim 750$ K). HATS-72 b, as the older and denser planetary twin (same mass, host star type and equilibrium temperature) to WASP-107 b, is expected to have stronger \methane features than its twin, implying WASP-107 b's hotter interior is the reason for its inflated radius \citep{Spake_2021}. The final analysis of these JWST observations has not yet been published. The host star HATS-72 also has no published carbon and oxygen abundances to date, so JWST's observations of HATS-72 b will rely on this work for additional context. 

\section{Observations \& Data} \label{sec:echoes_obs_data}
\subsection{High-Resolution, NIR IGRINS Spectra}
In this study, we use high-resolution $R \sim 45,000$, near-infrared (NIR) spectra obtained from the Immersion GRating INfrared Spectrograph (IGRINS, \citet{Park_2014, Mace_2018}) at the Gemini-South and Lowell observatories. IGRINS is a compact, cross-dispersed spectrograph offering simultaneous coverage of the H and K bands located between 1.45–2.45 microns.

Data were retrieved from the Raw \& Reduced IGRINS Spectral Archive \citep{Sawczynec_2025} including archival observations and data from our group's observing programs: GS-2023B-LP-108 (PI: Crossfield), GS-2024A-LP-108 (PI: Crossfield), and GS-2023A-Q-203 (PI: Crossfield). All IGRINS spectra are reduced using the IGRINS Pipeline Package V3 \citep{Sawczynec_2025}. This pipeline performs a number of reduction processes including flat-fielding, background removal, order extraction, distortion correction, wavelength calibration and telluric correction using A0V spectra to produce a final multi-order (23 orders in the H band, 21 orders in the K band) echelle spectrum per target. 

\begin{figure*}[h!]
    \centering
    \includegraphics[width=\textwidth]{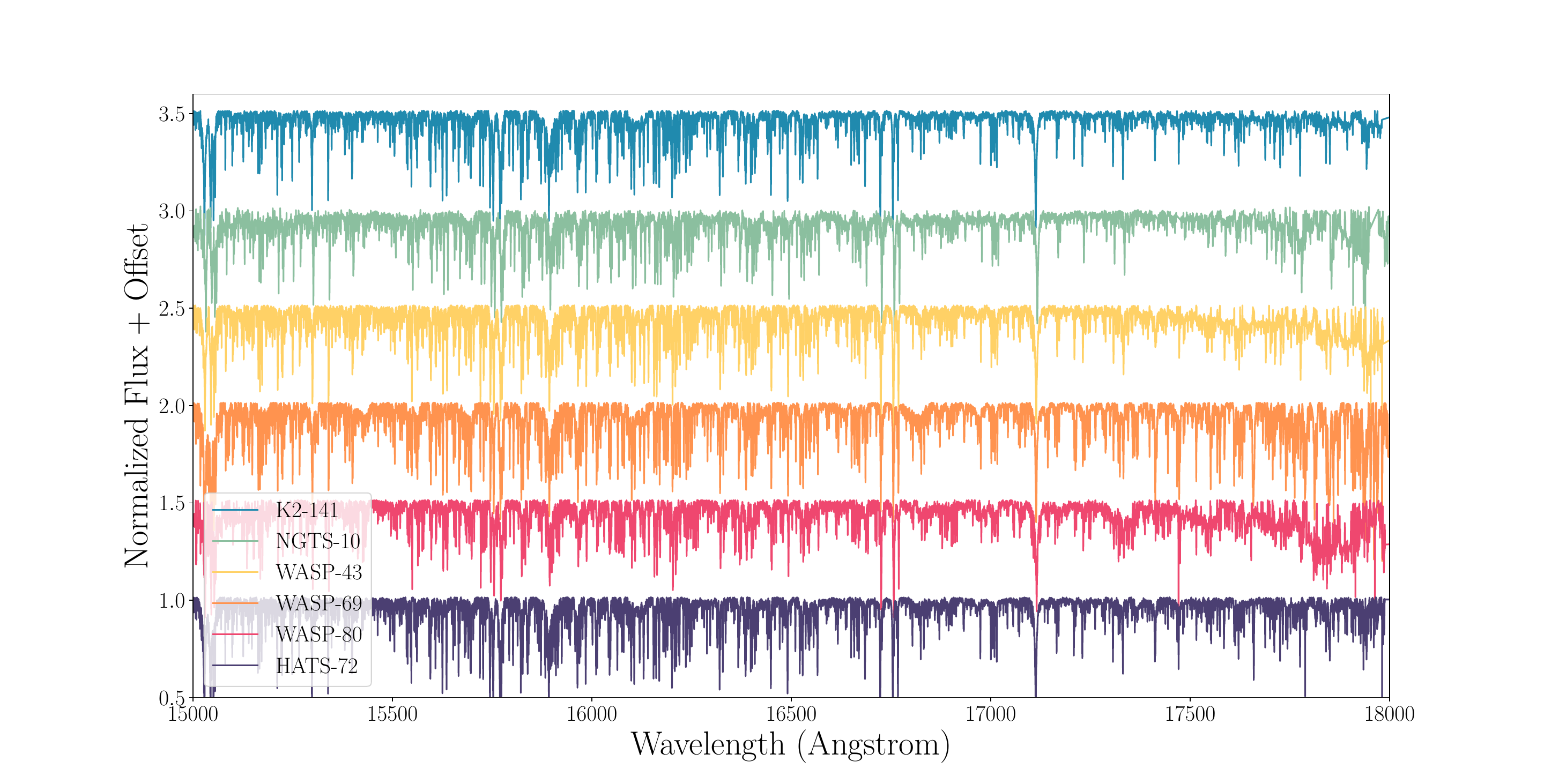}
    \includegraphics[width=\textwidth]{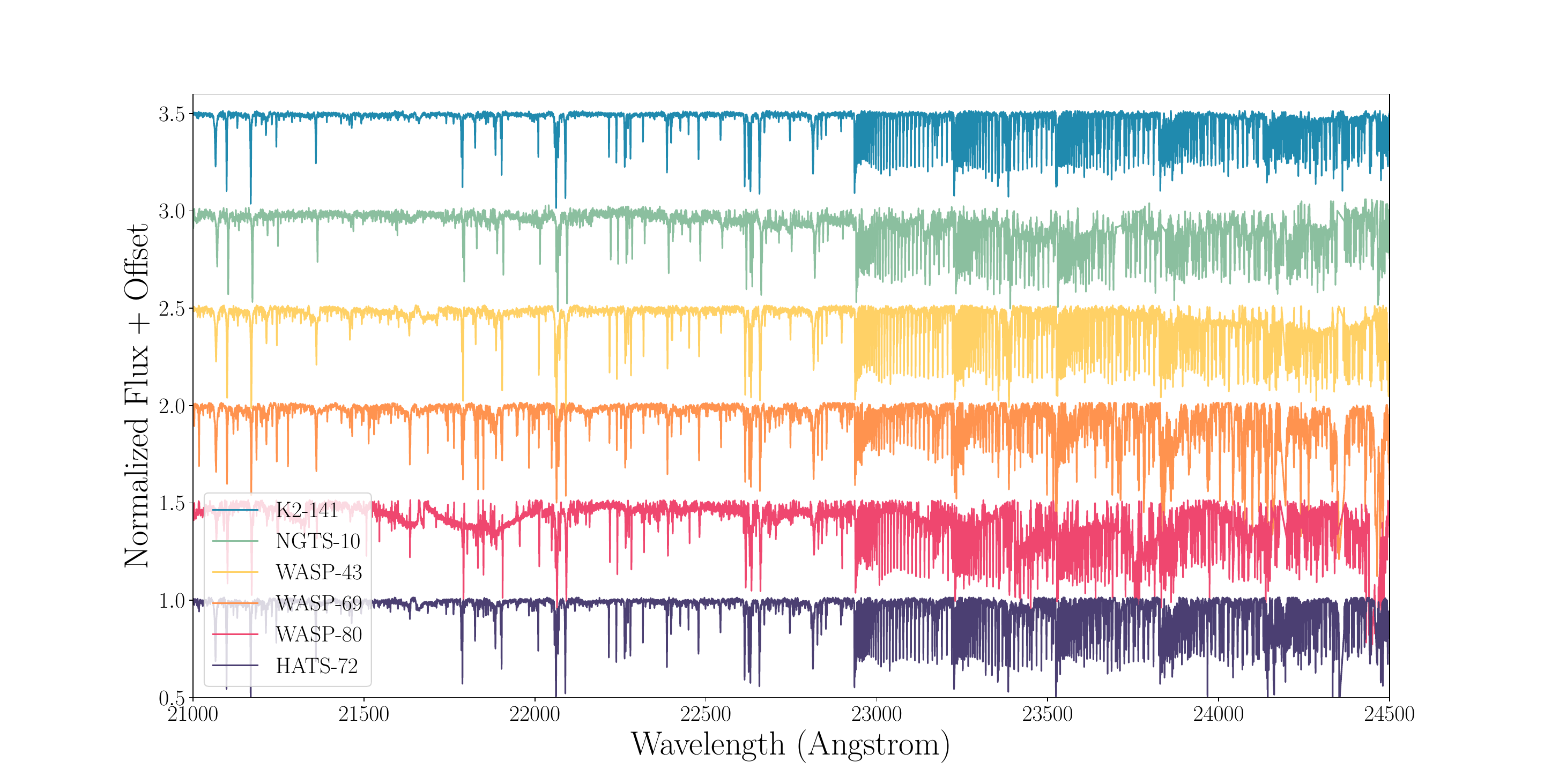}
    \caption{Plots of the observed, high signal-to-noise H (top panel) and K (bottom panel) band IGRINS spectra used in this abundance analysis. Oxygen abundances are determined from OH lines located in the H band (1.45–1.80 microns) while carbon, \crat, and \orat\ abundances are derived from various CO isotopologue lines located in the 2.3-2.45 micron region of the K band (1.95–2.45 microns).
    }
    \label{fig:kdwarf_waterfall}
\end{figure*}

The reduced multi-order spectra are then combined using the combspec tool in the SpeXTool package \citep{cushing:2004} and then flattened. These K dwarf spectra range in S/N from $S/N = 171.5$ in the H band and $S/N = 84.2$ in the K band for NGTS-10 to $S/N = 408.0$ in the H band and $S/N = 275.3$ in the K band for WASP-69. We plot these final, flattened spectra in Figure \ref{fig:kdwarf_waterfall}. Additional observational parameters are provided in Table \ref{tab:kdwarf_obs_param_tab}.

\begin{deluxetable*}{lcccccc}
\tablecaption{IGRINS Observational Parameters\label{tab:kdwarf_obs_param_tab}}
\tablewidth{0pt}
\tablehead{
\colhead{Parameter} &
\colhead{K2-141} &
\colhead{NGTS-10} &
\colhead{WASP-43} &
\colhead{WASP-69} &
\colhead{WASP-80} &
\colhead{HATS-72}
}
\startdata
Facility & Gemini South & Gemini South & Gemini South & Lowell Observatory & Lowell Observatory & Gemini South \\
Program ID & GS-2022B-Q-226 & GS-2023A-Q-203 & GS-2023A-FT-106 & --- & --- & GS-2023A-Q-203 \\
Date (UT) & 2022-09-06 & 2023-03-26 & 2023-04-12 & 2018-09-23 & 2017-09-12 & 2023-04-28 \\
Time (UT) & 02:12--02:15 & 00:43--02:12 & 00:30--00:33 & 04:04--04:06 & 07:06--07:11 & 08:52--08:57 \\
Integration Time (s) & 150 & 450 & 150 & 90 & 300 & 270 \\
Mean $H$-band S/N & 259.8 & 171.5 & 223.5 & 408.0 & 253.0 & 292.4 \\
Mean $K$-band S/N & 181.7 & 84.2 & 163.3 & 275.3 & 142.6 & 173.7 \\
\enddata

\tablenotetext{}{Observational parameters for the six K dwarfs observed with the IGRINS spectrograph \citep{Park_2014, Mace_2018}. Mean $H$-band S/N calculated over the 1.50--1.80~$\mu$m wavelength range. Mean $K$-band S/N calculated over the 2.20--2.45~$\mu$m wavelength range.}
\end{deluxetable*}

\section{Methods: Elemental Carbon and Oxygen Abundance Analysis}\label{sec:echoes_elemental_analysis}
The main goal of this study is to measure \crat\ and \orat\ ratios in a sample of K dwarf stars hosting high-priority giant exoplanets. Previous studies have made these measurements in solar type stars assuming solar carbon and oxygen elemental abundances \citep{Coria_2023, Coria_2024}. This is acceptable considering the stars in those samples are of solar metallicity; however, the same assumption cannot be made for the selected K dwarfs as most have non-solar[Fe/H] and a handful of which have non-solar [C/H] or [O/H]. The remaining K dwarfs have no [C/H] or [O/H] available in the literature at all. So, prior to determining \crat\ and \orat\ ratios, elemental carbon and oxygen abundances must be constrained. 

\subsection{Model Atmospheres, Line Data, and Spectral Synthesis}
The grids of model spectra for this study are generated following the methods described in \citet{Hejazi_2023} and \citet{Coria_2024}. We begin by synthesizing a grid of continuum-normalized model spectra using the TurboSpectrum 1D LTE radiative transfer code \citep{Plez_2012} together with 1D, hydrostatic, and plane-parallel MARCS atmosphere models, \citep{Gustafsson_2008} and a select set of atomic and molecular line lists presented in \citet{Hejazi_2023}. The line lists most crucial to a near-infrared carbon and oxygen abundance analysis are those for OH \citep{Goldman_1982, Goldman1998} and CO \citep{goorvitch:1994}.

First, we use the interpolation routine developed by T. Masseron to interpolate the MARCS model atmosphere associated with each star's physical parameters. These adopted T$_{eff}$, log g, and [Fe/H] (used interchangeably with [M/H], the bulk stellar metallicity relative to solar) values are listed in Table \ref{tab:stellar_abundances_Kdwarfs}. Then, we generate synthetic spectra using the interpolated model, while customizing the abundance of element X in Turbospectrum through an input quantity named ``relative abundance'' ABUND(X), which is related to the abundance [X/H], by

\begin{equation}
[\mathrm{X/H}] = \mathrm{ABUND(X)} + [\mathrm{M/H}],
\label{eq:abundance}
\end{equation}
where [M/H] indicates the overall metallicity. For carbon and oxygen, the abundances are given by
\begin{equation}
[\mathrm{C/H}] = \mathrm{ABUND(C)} + [\mathrm{M/H}],
\label{eq:carbon}
\end{equation}
and
\begin{equation}
[\mathrm{O/H}] = \mathrm{ABUND(O)} + [\mathrm{M/H}] + [\alpha/\mathrm{Fe}],
\label{eq:oxygen}
\end{equation}
respectively. Oxygen is an $\alpha$ element and is therefore subject to $\alpha$-enhancement scaling relations, with
$[\alpha/\mathrm{Fe}] = -0.4 \times [\mathrm{M/H}]$ for
$-1.0 \leq [\mathrm{M/H}] \leq 0.0$, and
$[\alpha/\mathrm{Fe}] = 0.0$ for $[\mathrm{M/H}] \geq 0.0$, as described in
\citet{Gustafsson_2008}. These corrections are negligible for our sample of stars with near-solar to super-solar metallicities.

The ``absolute abundance'' of element X is defined as
\begin{equation}
A(\mathrm{X}) =
\log_{10}\left(\frac{N_{\mathrm{X}}}{N_{\mathrm{H}}}\right) + 12,
\label{equ:abs_abund_def}
\end{equation}
where $N_{\mathrm{X}}$ and $N_{\mathrm{H}}$ are the number densities of
element X and hydrogen, respectively. Recalling that the abundance of
element X is defined relative to the Sun,
\begin{equation}
[\mathrm{X/H}]
=
\log_{10}\left(\frac{N_{\mathrm{X}}}{N_{\mathrm{H}}}\right)
-
\log_{10}\left(\frac{N_{\mathrm{X}}}{N_{\mathrm{H}}}\right)_{\sun},
\label{equ:abund_def}
\end{equation}
the absolute abundance can be expressed as
\begin{equation}
A(\mathrm{X})
=
A(\mathrm{X})_{\sun}
+
[\mathrm{X/H}].
\label{equ:abs_abund}
\end{equation}

The C/O ratio is then calculated from the absolute carbon and oxygen
abundances as
\begin{equation}
\mathrm{C/O}
=
\frac{N_{\mathrm{C}}}{N_{\mathrm{O}}}
=
10^{A(\mathrm{C})-A(\mathrm{O})},
\label{equ:abundance_ratio}
\end{equation}
with its propagated uncertainty, assuming independent uncertainties,
given by
\begin{equation}
\sigma_{\mathrm{C/O}}
=
\ln(10)\,(\mathrm{C/O})\,
\sqrt{
\sigma_{A(\mathrm{C})}^{2}
+
\sigma_{A(\mathrm{O})}^{2}
},
\label{equ:co_uncertainty}
\end{equation}
where $\sigma_{A(\mathrm{C})}$ and $\sigma_{A(\mathrm{O})}$ are the
uncertainties in the absolute carbon and oxygen abundances, respectively,
calculated as
\begin{equation}
\sigma_{A(\mathrm{C})}
=
\sqrt{
\sigma_{A(\mathrm{C})_{\sun}}^{2}
+
\sigma_{[\mathrm{C/H}]}^{2}
},
\end{equation}
and
\begin{equation}
\sigma_{A(\mathrm{O})}
=
\sqrt{
\sigma_{A(\mathrm{O})_{\sun}}^{2}
+
\sigma_{[\mathrm{O/H}]}^{2}
}.
\end{equation}

For each K dwarf, we generate a grid of synthetic spectra using its respective adopted T$_{eff}$, log g, and [Fe/H] values with different relative abundances of carbon and oxygen: ABUND(X) = +0.40, +0.30, +0.20, +0.15, +0.10, +0.05, +0.00, --0.10, --0.20. The selected K dwarfs have slightly sub-solar to super-solar [Fe/H] so the grids center on ABUND(X) = $+0.10$. 

\subsection{Line Identification and Continuum/Pseudo-continuum Placement}
We use a custom list of 27 CO lines and 78 OH lines to determine [C/H] and [O/H] abundances respectively. This list is adopted from \citet{Hejazi_2024} and includes line locations and the most appropriate continuum/pseudo-continuum regions around each line to be used for normalization. The observed spectrum must be normalized relative to the grid of synthetic spectra. This is done by fitting a low-order polynomial to the residuals R = O/S (where O is the observed flux and S is the interpolated synthetic flux at a given wavelength) of the data points within the selected continuum/pseudo-continuum regions. The normalized spectrum is obtained by dividing the observed flux by this final polynomial-fitted residual. The full continuum/pseudo-continuum placement procedure is described in \citet{Hejazi_2024, Hejazi2025}.


\subsection{Deriving Elemental Carbon and Oxygen Abundances}
\begin{figure*}[h!]
    \plottwo{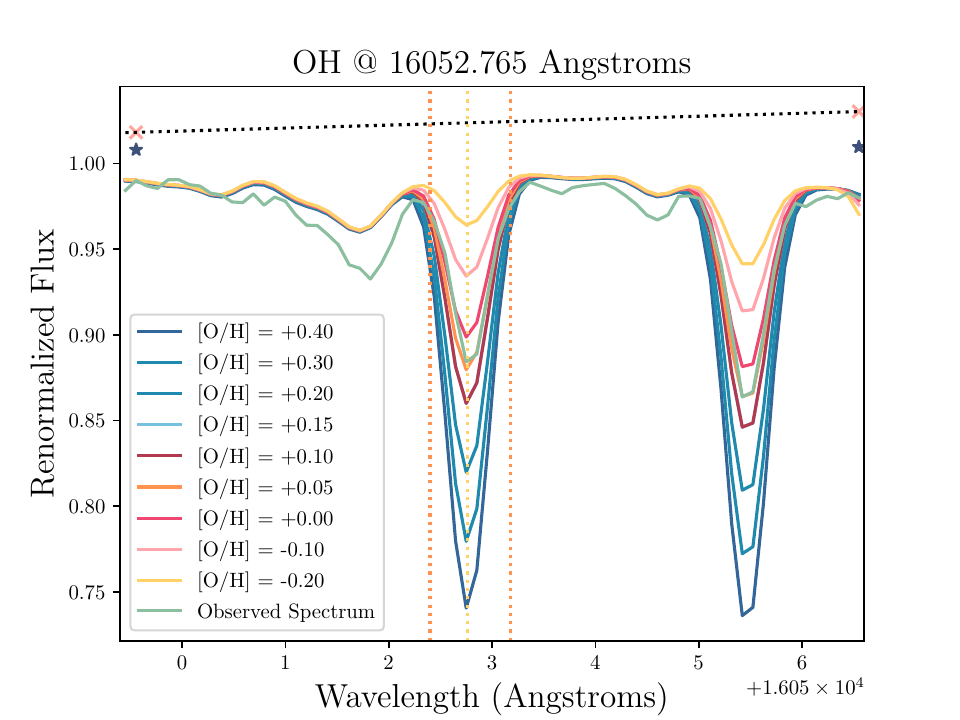}{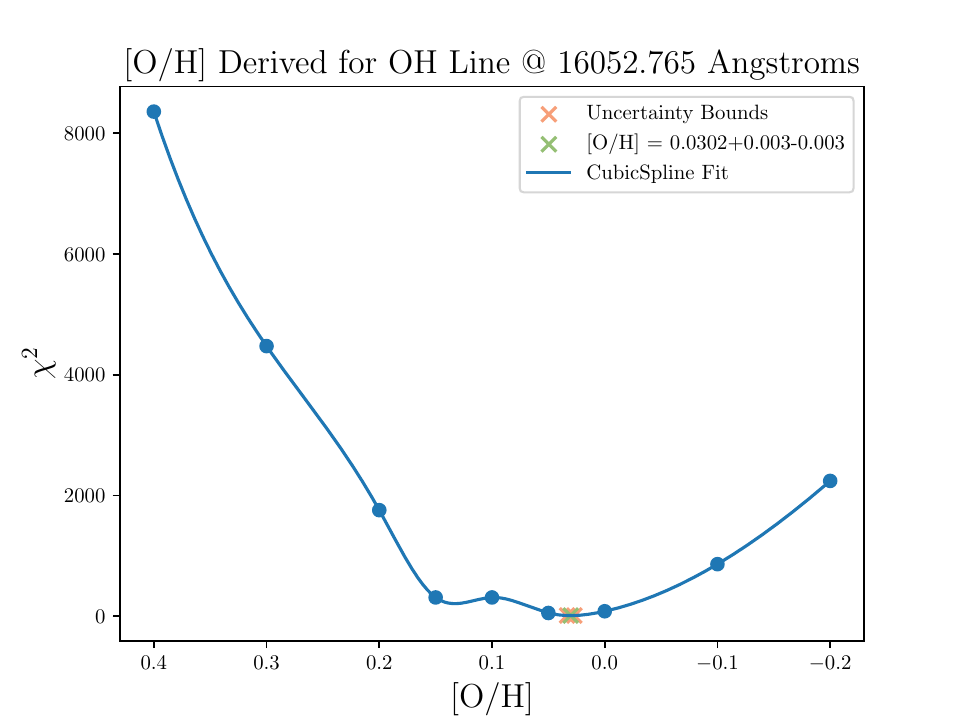}
    \plottwo{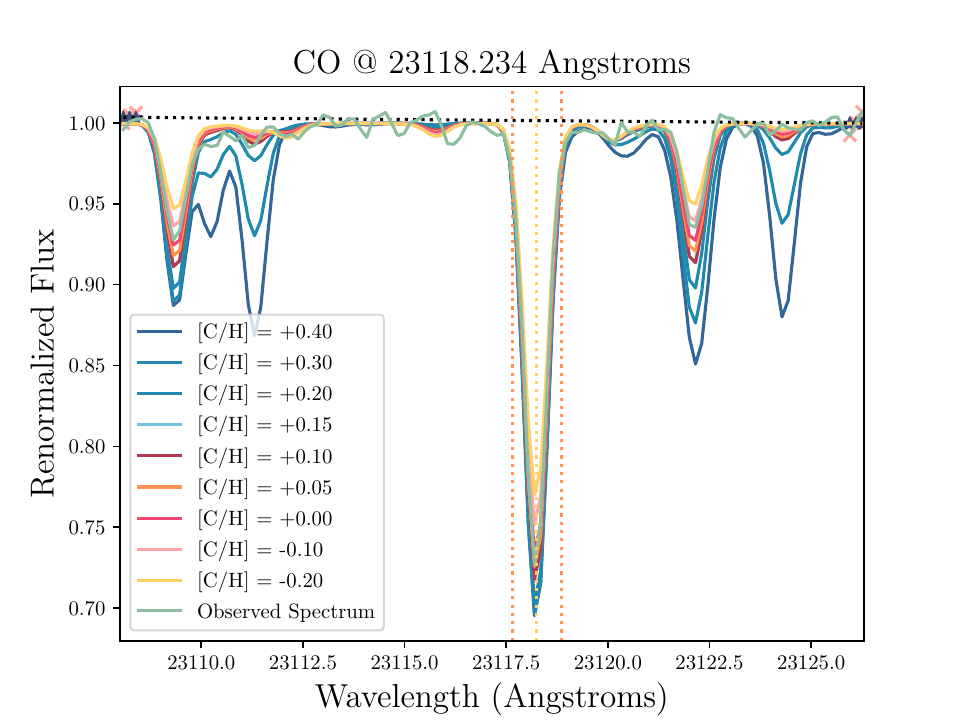}{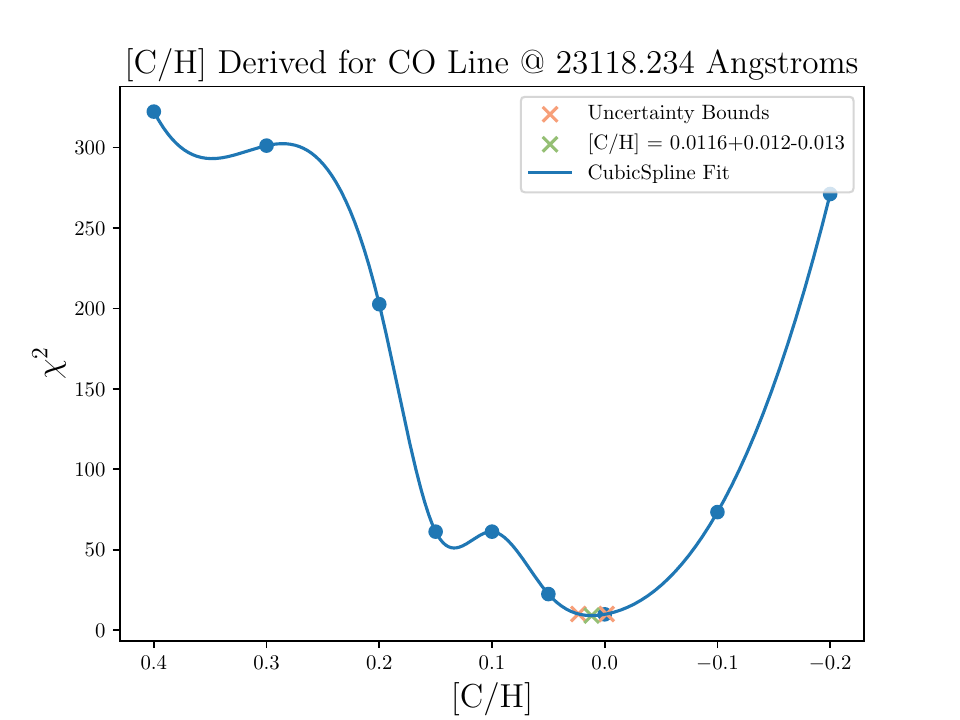}
    \plottwo{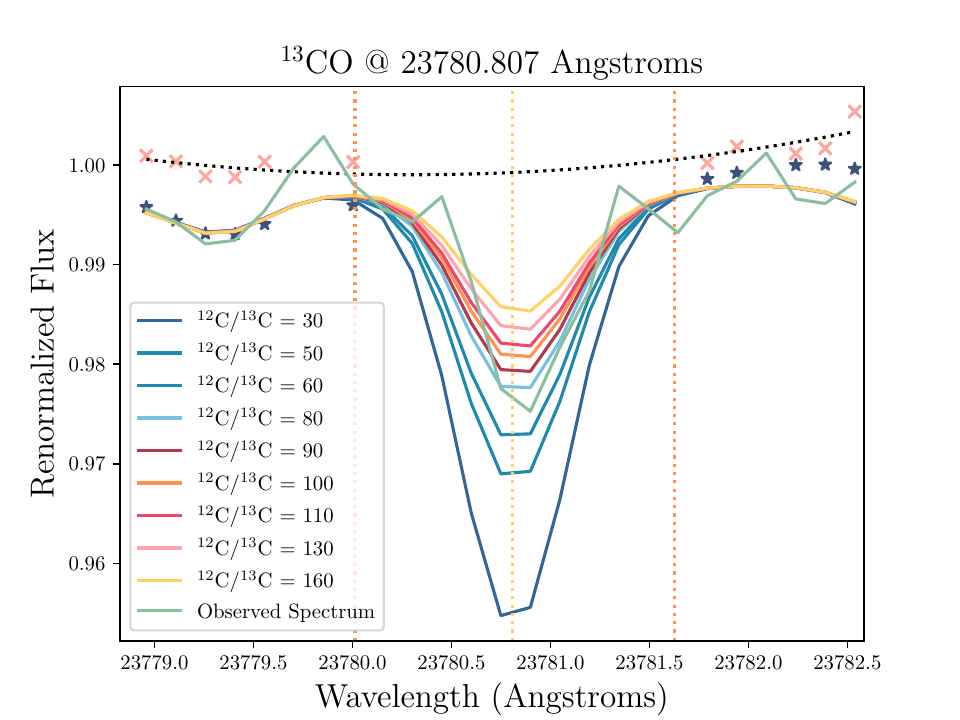}{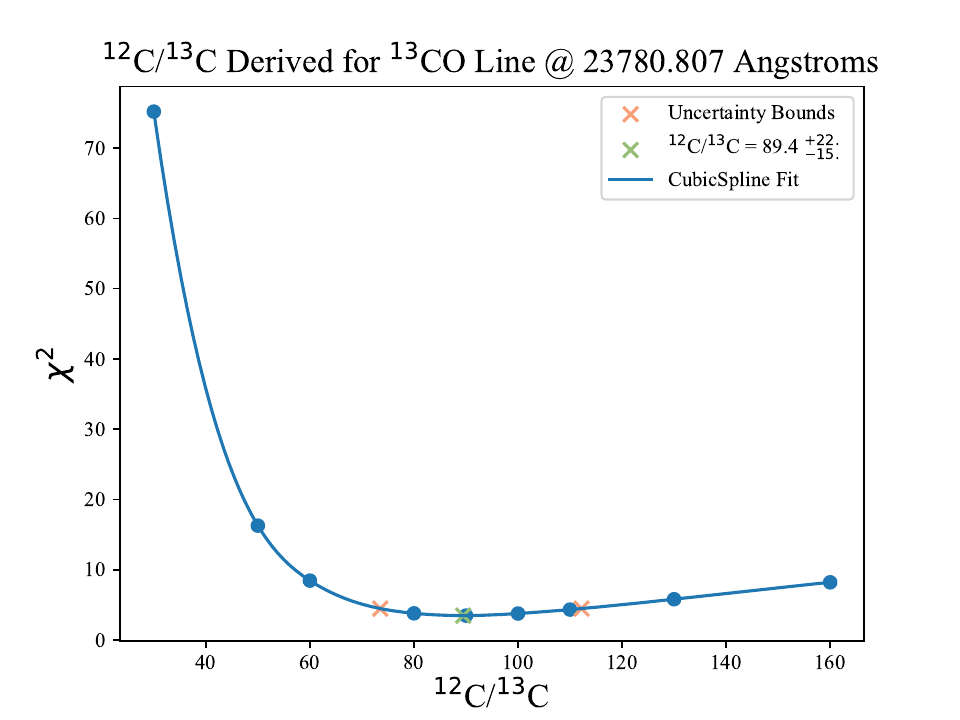}
    
    \caption{Deriving best-fit abundances for K2-141 using a $\chi^2$-minimization routine. $\chi^2$ is calculated between each synthetic spectrum corresponding to a different abundance value [X/H]. A cubic spline connects these $\chi^2$ points as shown in the right plots and is used to interpolate the value of [X/H] such that $\chi^2$ is minimized. This is the best-fit abundance value for a particular line. This routine is run only on the line core and includes only points inside the selected $\chi^2$ fit window for each line. Uncertainties represent 1$\sigma$ confidence intervals using the region where $\Delta\chi^2_{min} \leq 1$ \citep{avni:1976}. An example of this calculation is shown for an OH, CO, and $^{13}$CO line in the top, middle, and bottom rows respectively.
    }
   
    \label{echoes:chi_sqr_min}
\end{figure*}

Carbon [C/H] and oxygen [O/H] abundance values are determined line-by-line using a $\chi^2$-minimization routine as in \citet{Coria_2024}. A $\chi^2$ value is calculated between the normalized observed spectrum and each synthetic spectrum in the grid, where each synthetic spectrum corresponds to a different relative abundance value [ABUND(X) = +0.40, +0.30, +0.20, +0.15, +0.10, +0.05, +0.00, $-$0.10, $-$0.20] with the same physical stellar parameters. A cubic spline is fit to the set of $\chi^2$ and relative abundance points, and then we calculate the relative abundance where $\chi^2$ is minimized. This minimum in the $\chi^2$ curve represents the best-fit relative abundance value for the given line. This process is repeated for each OH and CO line; however, due to poor line-profile modeling, bad pixels in the observed spectrum, or issues with either normalization or fitting, not all lines are included in the final abundance calculation. An example of this process is shown in Figure \ref{echoes:chi_sqr_min}. 

A weighted average\footnote{Weights are given by $\frac{1}{\sigma_{stat}^2}$, the inverse-square of the statistical uncertainty on an individual [X/H] measurement.} of all the best-fit relative abundances for selected OH lines gives the relative oxygen abundance value for the star. Similarly, a weighted average of all the best-fit relative abundances for selected CO lines gives the relative carbon abundance value for the star. Final [O/H] and [C/H] abundances are obtained by adding the stellar metallicity values to the inferred relative abundances using Equations \ref{eq:carbon} and \ref{eq:oxygen}. 

The statistical uncertainties of these abundance measurements are given as the standard error on the mean of the abundances measured for each line. This way, we account for the overall spread of calculated abundance values and the number of lines used to derive the final abundance. Statistical uncertainties on the [C/H] values range from $\sim 0.01 - 0.03$ dex while those for [O/H] are $\sim 0.01$ dex.  \citet{Hejazi_2024} found that systematic uncertainties on these abundance values arise mostly from the choice of effective temperature and metallicity for the model spectra resulting in systematic uncertainties of $\sigma_{T} \leq 0.08$ and $\sigma_{Z} \leq 0.03$. We adopt these values here and report final uncertainties by adding $\sigma_{stat}$, $\sigma_{T}$, and $\sigma_{Z}$ in quadrature. Systematic uncertainties arising from the choice of surface gravity (log g) are negligible.

The final [C/H] and [O/H] values and uncertainties for each K dwarf in the sample, scaled according to Equations \ref{eq:carbon} and \ref{eq:oxygen}, are provided in the bottom panel of Table \ref{tab:stellar_abundances_Kdwarfs}. 

\section{Methods: Isotopic Carbon Abundance Analysis}\label{sec:echoes_isotopic_analysis}
To measure the \crat\ ratio for the the K dwarf sample, we follow the same methods as for deriving [C/H] and [O/H] described above. The MARCS model atmosphere grid is first interpolated to the adopted T$_{eff}$, log g, and [Fe/H] for each star as before. This time, the relative abundances for carbon and oxygen are fixed at the elemental abundance values derived above. This is done to ensure that the \crat\ model grids accurately reflect the carbon and oxygen ratios measured for each K dwarf rather than solar abundances.

Next, a grid of synthetic spectra is generated for each K dwarf using its respective adopted atmospheric parameters from Table \ref{tab:kdwarf_param_table}, the fixed relative abundances of carbon and oxygen, and varying \crat\ ratios: \crat\ = $30, 50, 60, 80, 90, 100, 110, 130, 160$. These \crat\ grids center around the solar value, \crat = $91.4 \pm 1.3$ \citep{ayres:2013}.

We employ the CO line list from \citet{goorvitch:1994}, which also contains information for minor isotopologues of CO, and TurboSpectrum's equivalent width calculation function \citep{Plez_2012} to identify strong lines that may be used to derive the \crat\ ratio. We visually inspect the isotopologue model grid at any locations where Turbospectrum calculates EQW $> 1\ m\AA$. This filtering leaves thirteen $^{13}$CO regions for which we determine suitable continuum/pseudo-continuum regions for normalization and $\chi^2$ fitting windows as in \citep{Hejazi_2023, Coria_2024}. These normalization and fitting regions are available in the accompanying machine-readable line list (see Table \ref{tab:a1}). It is important to note that the line centers and $\chi^2$ fit windows provided for the minor isotopologue \cmin O may, but do not necessarily refer to individual lines and can refer to regions containing multiple blended lines. Furthermore, the \crat\ ratios derived from these overtone rovibrational band lines are highly sensitive to the normalization parameters, requiring a careful continuum/pseudo-continuum selection, but also to spectral noise and the choice of fundamental stellar parameters for the model spectra. For this reason, we consider more conservative uncertainties that account for the systematics associated with these measurements in addition to the relatively small statistical uncertainties as in \citet{Coria_2024}. In the previous study, $\sigma_{sys} = 6-7$ for high quality VLT/ESPRESSO and Keck/HIRES spectra from a well-characterized star (WASP-77 A), but since our spectra are lower S/N and our uncertainties on the stellar effective temperatures are higher, we adopt $\sigma_{sys} = 10$. Final uncertainties are estimated by adding $\sigma_{stat}$ and $\sigma_{sys}$ in quadrature. 

Although the noise in WASP-80's spectrum prevents a reliable \crat\ measurement, we are able to derive \crat\ ratios for the remaining five targets using at least three \cmin O lines. These values are all sub-solar ranging from \crat\ $= 38 - 52$, indicating some \cmin\ enrichment for all five hosts.

Even using the high signal-to-noise IGRINS spectra ($84 < $ S/N $ < 275$) poses a challenge for deriving \crat\ ratios from weak \cmin O isotopologue lines. The C\omin\ isotopologue lines are significantly weaker, so the prospects for deriving reliable stellar \orat\ ratios from K band spectra are poor as shown in \citet{Gonzalez_Picos_2025}. This means that perhaps deriving \crat\ and \orat\ ratios is best done in the M band using \textit{fundamental} rovibrational CO isotopologue lines as in \citet{crossfield:2019a}, \citet{Coria_2023}, and \citet{Gonzalez_Picos_2026_TRAPPIST1}.

\section{Discussion}\label{sec:discussion}
\subsection{Stellar Activity Effects}
Targets NGTS-10, WASP-43, WASP-69, and WASP-80 are known to have high activity indexes, but this does not significantly affect our abundance measurements. Magnetic effects are negligible for molecular OH and CO lines used in this work \citep{Reiners_2012, Cristofari_2023} as lines from the first overtone of CO, and similarly the NIR OH band, form deeper in the star and are significantly less affected by any chromospheric activity \citep{Ayres_1989, Wiedemann_1994}. WASP-69 is the most active star in our sample and known to exhibit starspots and faculae \citep{Petit_dit_de_la_Roche_2024}. However, the effect is minimal: TESS light curve analysis yields a spot filling factor of 0.77 of the stellar surface \citep{Chakraborty_2024}. For a photospheric temperature of approximately 4700 K and spots several hundred Kelvin cooler, this filling factor produces only a few-kelvin change in the flux-weighted effective temperature, suggesting that the effect of the observed spot activity on our derived abundances is small compared with the current uncertainties.

\subsection{The Stellar Age Problem}
While many fundamental stellar parameters like mass, radius, and chemical composition can be measured with relative ease, determining the age of an individual star remains notoriously difficult. This is a problem because so many areas of astronomy rely on this chronological context for testing stellar evolution models, understanding galactic archaeology and evolution, and exploring the formation and evolution of exoplanetary systems. Regarding stellar chemical abundances, stellar ages enable the search for abundance-age relationships for different groups of stars. Because solar twin stars are the most amenable to chemical abundance measurements, several relationships have been found between the isochrone ages of these stars and individual elemental abundance ratios, establishing abundance ratios as stellar age proxies. We refer the reader to Section 2.3 of \citet{Coria_2023} and references therein for more details on chemical abundances as stellar chronometers. These abundance-age relationships break down, however, beyond solar twin parameters and remain largely unexplored due to the lack of chemical abundance measurements-- and associated challenges in making them-- for cooler K/M dwarfs as well as the difficulty in constraining the ages for these stars.

The standard approach for determining stellar ages is to apply isochrones (curves on a Hertzsprung-Russell diagram that connect stars of the same age but different initial masses) that relate the star's physical properties (e.g. effective temperature, luminosity, chemical composition) to stellar evolution models. Notable isochrone packages in astronomy include Yonei-Yale isochrones \citep{Yi_2001, Kim_2002, Yi_2003, Demarque_2004}, the PAdova and TRieste Stellar Evolution Code (PARSEC) \citep{Bressan_2012, Nguyen_2022}, and MESA Isochrones and Stellar Tracks (MIST) \citep{Choi_2016, Dotter_2026}. However, this approach is notoriously challenging for cool dwarfs which move imperceptibly across this parameter space while on the main sequence \citep{veyette:2018, Angus_2019, Choi_2016, Dotter_2026}. Furthermore, isochrones are too closely spaced in effective temperature vs. luminosity space for field-age main-sequence cool dwarfs, meaning isochrone fitting alone often produces unrealistic and unreliable ages for cool dwarf stars.

One alternative is to instead use asteroseismology to measure surface oscillations and infer a star's evolutionary stage based on its internal structure and composition. Stellar surface oscillations reveal internal structure allowing for precise determination of  radii, masses, and ages \citep{Lebreton_2009, Garcia_Ballot_2019}. This again is challenging for cool dwarfs because the oscillations are so weak, and the technique has only been applied to a handful of bright, nearby K dwarfs like $\epsilon$ Indi, HD 219134, and $\sigma$ Draconis \citep{Campante_2024, Hon_2024, Li_2025}.

Other approaches for determining stellar ages rely on lithium abundances or stellar activity indicators. Lithium is a powerful stellar age indicator because stars destroy their initial lithium supply as they age through mixing and burning processes. This results in an observable Li depletion by a factor of $\sim$50 as age increases from 1 to 9 Gyr \citep{Carlos_2016}. However, this Li clock is most sensitive for younger stars with ages $< 1$ Gyr and provides the best age constraints for coeval groups of stars versus single stars \citep{Jeffries_2014}. Regarding stellar activity, there is a strong anticorrelation between chromospheric activity and age leading to the use of the Ca II H and K emission index as an age indicator for late F- to early K-type stars \citep{Mamajek_Hillenbrand_2008}. Other relations including the H-alpha emission index can only be used to roughly estimate ages of field stars with 4000 K $<$ T$_{eff}$ $<$ 5500 K with an accuracy of about 40-60 percent \citep{Zhang_2019}. In addition, these chromospheric activity age indicators are best suited to stars younger than 2 Gyr \citep{Pace_2013}.

The last ray of hope for obtaining cool dwarf ages is through gyrochronology which assigns ages based on empirical temperature-rotation-age relations observed in clusters. Gyrochronology, in contrast to the methods described above, is ideal for assigning ages to single main sequence stars, middle-aged/old stars, and cool K/M dwarfs \citep{Gaidos_2023, Bouma_2024, Lu_2024}. Stellar ages have been primarily estimated through gyrochronology for many planetary systems, particularly for those discovered by the WASP/SWASP surveys \citep{Bonomo_2017}. In most cases, gyrochronology- and isochrone-based stellar ages agree closely \citep{Bonomo_2015}, but gyrochronological ages may not be as reliable for exoplanet hosts \citep{Angus_2015} because of possible spin-up of the host stars by their close-in exoplanets \citep{Pont_2009, Maxted_2015}. However, they are often the easiest to calculate for cool dwarf stars as many often have reliable and readily accessible rotation periods from WASP, Kepler, or TESS light curves \citep{Hellier_2011, Bouma_2024, Boyle_2026} in addition to other required fundamental parameters like temperature, distance, and magnitude.

\subsection{Literature Ages for the K Dwarf Sample}
Literature ages for late K/M dwarfs are exceedingly rare, however, given that the planetary systems in this work are prime targets for ground- and space-based characterization, several efforts have been made to estimate the host star ages. Table \ref{tab:ages_table} provides a comprehensive list of the stellar rotation periods and ages, both isochrone and gyrochrone, available for our K dwarf sample: K2-141, NGTS-10, WASP-43, WASP-69, WASP-80, and HATS-72. Most have at least one published rotation period, isochrone age, and gyrochrone age (HATS-72 lacks a gyrochrone age). The better studied WASP systems have multiple and sometimes conflicting reported values for each of these parameters. One example is the conflicting isochrone age estimates for WASP-43 $Age = 0.2 \pm 0.6$ Gyr \citep{Bonfanti_2015} or $Age = 11.21 \pm 1.5$  Gyr\citep{Huang_2022}. Similarly, WASP-80's rotation period analyses show a concerning range of viable periods from 3 to 23 days \citep{Salz_2015, Penev_2018, Triantafillides_2026}. Some values in the table, like WASP-69's $1.8 \pm 0.8$ Gyr isochrone age \citep{Allart_2025}, are actually composite ages derived from several of the numerous techniques presented in the previous section. \citet{Allart_2025} uses PARSEC isochrones, Li abundances, Ca II H\&K emission, chromospheric age, gyrochronology relations, and metallicity-age relations to determine their final age value for WASP-69. Similarly, WASP-69's composite gyrochrone age of $1.1 \pm 0.15$ Gyr is derived using some older rotation-age relationships \citep{Barnes_2007}, Li abundance-age relations \citep{Sestito_Randich_2005}, and activity-age relationships\citep{Mamajek_Hillenbrand_2008}. 

\subsection{Determining Isochrone \& Gyrochrone Ages for the K Dwarf Sample}
Because there are significant differences in the methods and resulting literature ages for our targets, we decide to conduct a new isochrone fitting and gyrochrone age analysis. 

We use the spectroscopic parameters listed in Table \ref{tab:kdwarf_param_table} along with the Gaia DR3 parallax and magnitudes ($G$, $B_P$, and $R_P$), Two Micron All Sky Survey (2MASS) magnitudes ($J$, $H$, and $K_S$), and Wide-field Infrared Survey Explorer (WISE) magnitudes ($W_1$, $W_2$, and $W_3$) to perform an isochrone fit in order to derive the physical parameters of the host star, including the age. The spectroscopic parameters, parallax, and magnitudes are used as priors to determine the goodness of fit. We use the \texttt{isochrone} package \citep{Morton_2015} to generate the MIST (MESA Isochrones and Stellar Tracks) isochrone models \citep{Choi_2016} used to sample the stellar parameters and find the best-fit parameters by using a Markov Chain Monte Carlo (MCMC) routine using the \texttt{emcee} package \citep{ForemanMackey_2013}. We obtain ages consistent with our upper bound of 14 Gyr for 5 of our 6 targets, with NGTS-10 being the exception with an age of $10.1^{+1.7}_{-1.8}$ Gyr. The issue of K-dwarf isochrone ages greater than the age of the universe is not new, particularly in the case of WASP-69 \citep{Maxted_2015}. These isochrone ages are listed at the bottom of Table \ref{tab:ages_table}.

We use a simple code, \texttt{GPGyro} to infer gyrochronological ages of single, main-sequence stars between 0.67 and 14 Gyr from effective temperature, rotation period measurements, absolute magnitude and parallax \citep{Lu_2024}. Effective temperature, magnitude, and parallax values for the sample are found in Table \ref{tab:kdwarf_param_table}. Rotation periods are listed in Table \ref{tab:ages_table}. The effective temperature, rotation period, and absolute magnitudes are passed to \texttt{GPGyro} which calculates a gyrochrone age. For the targets with multiple, consistent rotation periods, we perform the gyrochrone age calculation using a weighted average of the stellar rotation periods. WASP-80 notably has three conflicting rotation periods $8.1 \pm 0.8$ \citep{Salz_2015}, $3.15 \pm 0.01$ \citep{Triantafillides_2026}, and $23.5 \pm 3.0$ \citep{Penev_2018} days derived from WASP light curves, TESS light curves, and stellar-spin models respectively. We consider only the rotation periods derived from WASP-80 light curve analyses for uniformity. Because the two light-curve-derived rotation periods are quite different, we run \texttt{GPGyro} separately for each rotation period for this target. 

\begin{figure*}[h!] 
    \includegraphics[width=1.0\textwidth]{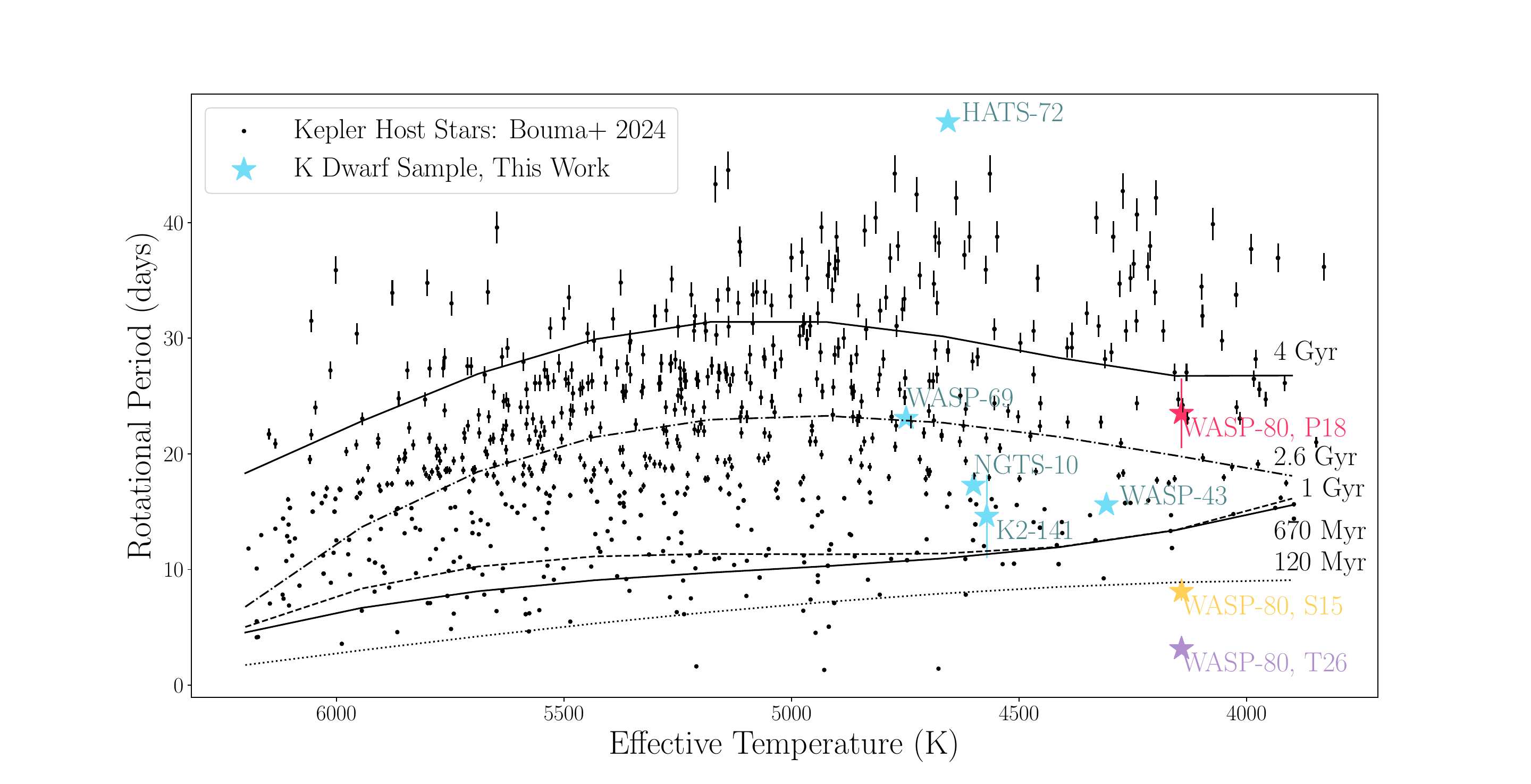}
    \includegraphics[width=1.0\textwidth]{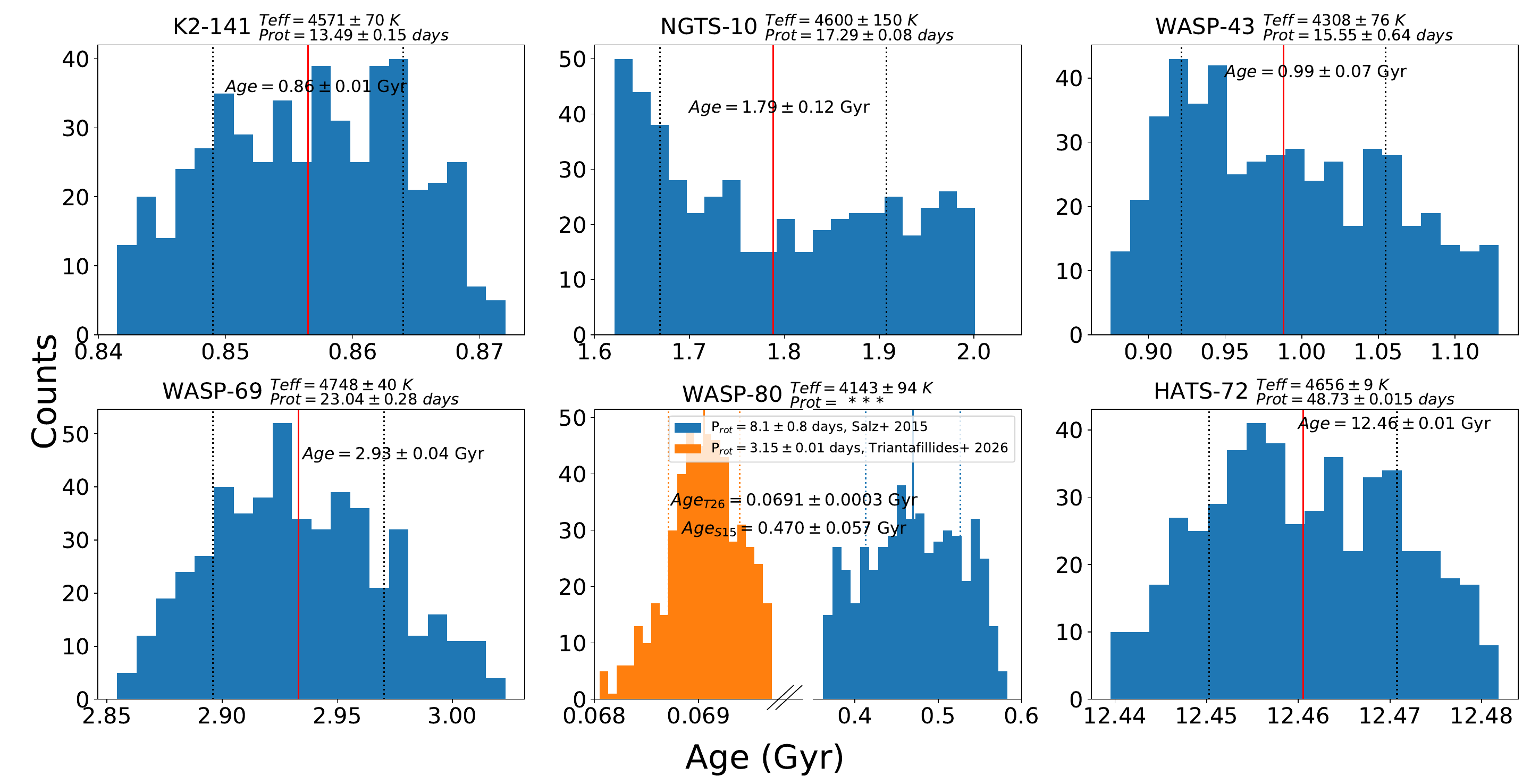}
    
    \caption{Comparison to Kepler Field gyrochrones and \texttt{GPGyro} Age Distributions. \textbf{Top:} This stellar rotation period vs. effective temperature plot shows gyrochrone curves determined for Kepler field stars \citep{Bouma_2024}. The K dwarfs analyzed in this work are overplotted to demonstrate that the adopted effective temperatures and rotation periods mostly agree with existing gyrochrones in the literature except HATS-72 which lies outside the explored parameter space and WASP-80 which appears younger on this plot than the derived gyrochrone ages suggest. \textbf{Bottom:} Histograms representing Monte Carlo realizations of 500 \texttt{GPGyro} \citep{Lu_2024} runs per target. The mean \textbf{(solid lines)} and standard deviation \textbf{(dotted lines)} of each distribution are adopted as the final gyrochrone age and uncertainty.
    }
    \label{fig:gyrochrones_age_dist}
\end{figure*}

Following the methods in \citet{Gaidos_2023}, we perform Monte Carlo realizations (500 points) of these calculations to estimate the uncertainty in the age, incorporating errors in rotation period, effective temperature, and the intrinsic gyrochronology relations. Histograms for the resulting age distributions are shown in Figure \ref{fig:gyrochrones_age_dist} and the mean and standard deviation of the distributions are adopted as the final age and uncertainty as listed at the bottom of Table \ref{tab:ages_table}.

The new gyrochrone ages for our K dwarf sample are mostly consistent with previous gyrochrone age estimates within the uncertainties. Figure \ref{fig:gyrochrones_age_dist} (top) shows the rotation period vs. effective temperature gyrochrone curves from \citet{Bouma_2024} used to determine rotation-based ages for stars in the Kepler field younger than 4 Gyr. HATS-72's slow rotation period is inconsistent with any of the gyrochrones, indicating an age $> 4$ Gyr. WASP-69 lies right along the 2.6 Gyr gyrochrone, and this is consistent with the $2.86 \pm 0.38$ Gyr \texttt{GPGyro} age derived in this work. K2-141, NGTS-10, and WASP-43 lie between the 2.6 and 1 Gyr gyrochrones, again consistent with our new estimates. Finally, our new age estimates for WASP-80, 69 Myr and 470 Myr based on the \citet{Triantafillides_2026} and \citet{Salz_2015} rotation periods respectively, both agree and disagree with \citet{Bouma_2024} gyrochrones. The former age lies well below the 120 Myr isochrone and is consistent with a $< 120$ Myr age. The latter age lies along the 120 Myr gyrochrone, but is several hundred Myr higher, in contention.

Compared to previous isochrone age estimates, the K dwarfs appear significantly younger according to rotation-age relationships with K2-141, NGTS-10, WASP-43, and WASP-80 having ages $< 2$ Gyr. The exceptions are WASP-69 which seems to be more of a middle-aged star compared to previous gyrochrone age estimates and HATS-72 which appears to be old $\sim 12$ Gyr regardless of the isochrone or gyrochrone age estimate.

\subsection{Galactic Chemical Evolution Context \& Tentative \crat\ Trends}
Now with chemical abundances and stellar ages in hand, we may use Galactic chemical evolution (GCE) models to contextualize our results and search our K dwarf sample for abundance-metallicity or abundance-age trends. In this study, we consider various state-of-the-art GCE models \citep{Romano:2017, prantzos:2018, kobayashi:2020} which have previously issued a call for more observational constraints from cool dwarf stars. We refer the reader to \citet{Coria_2023}, Section 5, for a summary of these GCE models and a detailed description of their different nucleosynthetic prescriptions.

\begin{figure*}[h!] 
    \includegraphics[width=1.0\textwidth]{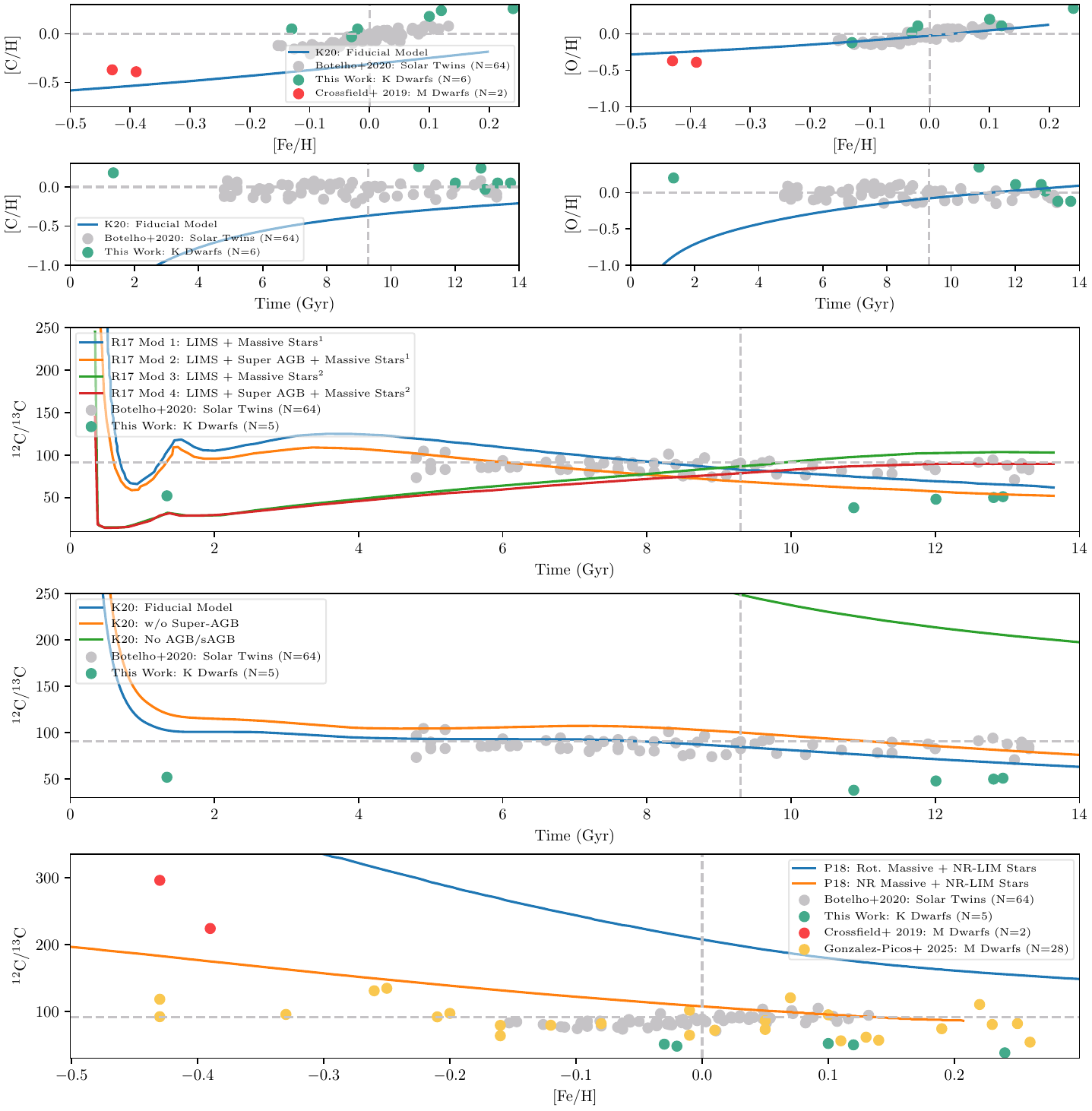}
    
    \caption{Elemental and Isotopic Abundance Comparison to Galactic Chemical Evolution (GCE) Models. \textbf{First Row:} Elemental carbon and oxygen abundances vs. [Fe/H] from our K dwarf sample, solar twins \citep{Botelho_2020}, and M dwarfs \citep{crossfield:2019a}, plotted against the fiducial GCE model from \citet{kobayashi:2020}. \textbf{Second Row:} Same as first row, but now using time on the x-axis instead of [Fe/H]. The time axis represents the age of the universe, meaning any stellar abundances are plotted at their expected formation times. \textbf{Third Row:} \crat\ ratios derived in this work and the \citet{Botelho_2020} solar twin sample plotted against four GCE models from \citet{Romano:2017}, each with a different nucleosynthetic prescription. \textbf{Fourth Row:} Same as previous, but now using three GCE models from \citet{kobayashi:2020}. \textbf{Fifth Row:} Similar to previous, but now using two GCE models from \citet{prantzos:2018} and [Fe/H] replacing time on the x-axis. The cool dwarfs from this work, \citet{Botelho_2020, crossfield:2019a, Gonzalez_Picos_2025} are added back in as they all have [Fe/H] measurements, but not ages. In all the plots above, solar values are marked by gray dotted lines. We refer the reader to \citet{Coria_2023}, Section 5, for a summary of these GCE models and a detailed description of their different nucleosynthetic prescriptions.
    }
    \label{fig:GCE_mod}
\end{figure*}

\begin{figure*}[h!]
    \includegraphics[width=1.0\textwidth]{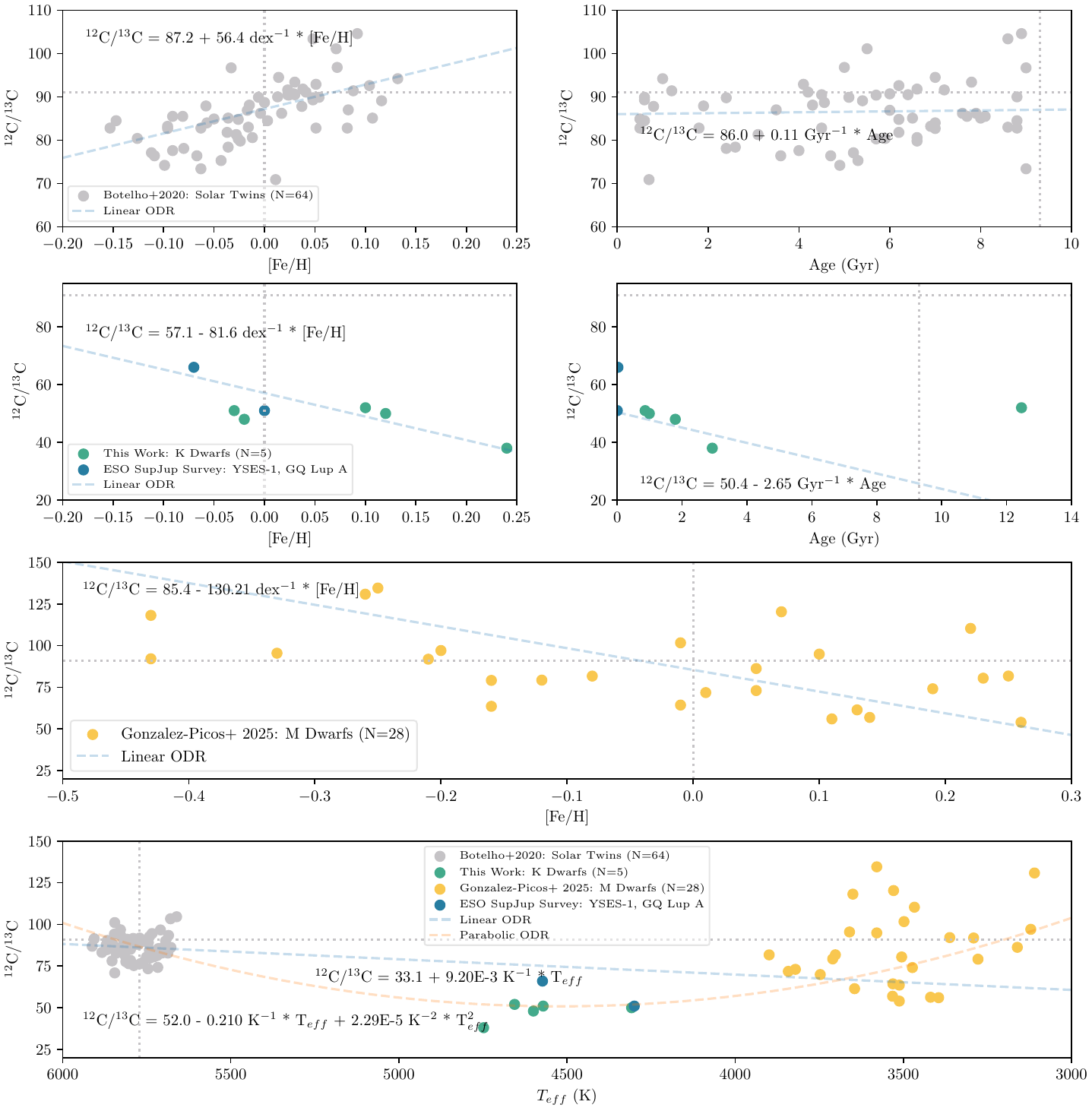}
    
    \caption{Searching for abundance-metallicity-age trends. \textbf{First Row:} \crat\ vs. [Fe/H] (left) and Stellar Age (right) for \citet{Botelho_2020} solar twin sample. \textbf{Second Row:} Same as above, but now for K dwarfs analyzed in this work and in \citet{Zhang_2024, Gonzalez_Picos_2025_GQLup}. \textbf{Third Row:} Similar to above, but for M dwarfs that only have [Fe/H] measurements, not ages. \textbf{Fourth Row:} \crat\ vs. effective temperature for all cool dwarfs plotted above. In all plots, dotted lines represent solar values while dashed lines represent the corresponding linear or parabolic orthogonal distance regression (ODR) fits, each labeled with its corresponding equation of the \crat as a function of effective temperature.
    }
     \label{fig:crat_trends}
\end{figure*}

In Figure \ref{fig:GCE_mod}, we plot several Galactic chemical evolution models \citep{Romano:2017, prantzos:2018, kobayashi:2020} against [C/H], [O/H], and \crat\ ratios from several cool dwarf stars \citep{crossfield:2019a, Botelho_2020, Gonzalez_Picos_2025} including the K dwarf sample from this work. Regarding elemental carbon and oxygen abundances (top two rows in Figure \ref{fig:GCE_mod}), the younger members of our K dwarf sample appear to follow trends previously observed in solar twin stars from \citet{Botelho_2020}. When looking at the K20 fiducial model, [O/H] is better reproduced than [C/H]. While O and Fe which have single production mechanisms (core-collapse supernovae and Type Ia supernovae, respectively), carbon is produced via two major channels: core collapse supernovae and low- to intermediate-mass asymptotic giant branch (AGB) stars. This makes [C/H] more difficult to reproduce using first-principle GCE models like Kobayashi+ 2020 which do not consider carbon yields from fast-rotating massive stars. Moreover, the dominant carbon-producing channel is still heavily debated as discussed in \citet{Boyea_2026}.

The \crat\ for our one old K dwarf, HATS-72, helps us better understand chemical evolution in the Galaxy's early history-- favoring Models 1 and 2 over Models 3 and 4 from \citet{Romano:2017} like the rest of our K dwarf sample as seen in Figure \ref{fig:GCE_mod}. In the \citet{kobayashi:2020} GCE model plot from Figure \ref{fig:GCE_mod}, Model 3 without nucleosynthetic contributions from AGB and super-AGB stars significantly underestimates the \cmin\ -enrichment observed in several cool dwarf stars which instead favor Models 1 and 2. Similarly, most cool dwarf \crat\ measurements favor \citet{prantzos:2018} Model 2 with nucleosynthetic prescriptions from non-rotating massive stars and non-rotating low-to-intermediate mass (NR-LIM) stars rather than Model 1 which instead considers rotating massive stars and NR-LIM stars. 

From Figure \ref{fig:GCE_mod}, it is clear that \crat\ GCE models still need more observational constraints from cool dwarf stars, but particularly from older, low metallicity stars.

In Figure \ref{fig:crat_trends}, we search the observed \crat\ ratios for abundance-metallicity-age trends. Each sample of stars is fit using a linear orthogonal distance regression (ODR) as there are uncertainties on both the x- (i.e. \crat) and y-values (i.e. [Fe/H], stellar ages, and effective temperatures) considered. \textit{First Row:} While there is a tentative trend between \crat\ and [Fe/H], there is no clear trend between \crat\ and the isochrone ages for solar twin stars like those in \citet{Botelho_2020}. This is inconsistent with most GCE models which predict \cmin-enrichment with increasing [Fe/H] and younger ages \citep{prantzos:2018, kobayashi:2020}. The exceptions are Models 3 and 4 from \citet{Romano:2017} which do predict increasing \crat\ ratios for younger stars. Only the K dwarfs ``\crat vs. Age (Gyr)” plot (Figure \ref{fig:crat_trends}, second row, right panel) is consistent with these model trends as the \crat ratio increases towards the younger stars in the sample. Because our K dwarfs fit the trend but not the \crat values predicted by Models 3 and 4 (Figure \ref{fig:GCE_mod}, third row), this could be indicative of inaccurate nucleosynthetic yields at earlier times in the galaxy’s history. More observational constraints could help address this inconsistency, but the \crat ratio parameter space for middle-aged dwarf stars remains largely unexplored. \textit{Second Row:} For the K dwarf from this work and the ESO SupJup Survey \citep{Zhang_2024, Gonzalez_Picos_2025_GQLup} there appears to again be a trend between \crat\ and [Fe/H] but now opposite to that observed for solar twins (left). There also appears to be a tentative trend with age (right), but the lack of constraints from older stars makes this unreliable. While they follow the general trends predicted by the GCE models considered here, the models somewhat underestimate the \cmin-enrichment observed in these stars. \textit{Third Row:} This M dwarf sample from \citet{Gonzalez_Picos_2025} is much larger than the K dwarf sample in this work. While the observed \crat\ ratios hover around the solar value, the scatter is much more pronounced for these M dwarfs than for the solar twins from \citet{Botelho_2020}. Nevertheless, the observed trend appears consistent with that observed for K dwarfs and most GCE models considered here. \textit{Fourth Row:} We plot \crat vs. effective temperature for the stellar samples discussed above. A parabolic orthogonal distance regression fits the data well; however, the observed trend of K dwarfs leaning more \cmin-rich than either the solar twins or M dwarfs may not be due to their unique range of effective temperatures but rather Galactic chemical evolution as the K dwarf sample is likely younger than either the hotter or cooler stars. A couple caveats to encourage further exploration: (1) the M dwarfs from \citet{Gonzalez_Picos_2025} do not have measured ages which could alter our interpretation; (2) our K dwarf sample is quite small and does not probe the middle-aged (4-10 Gyr) parameter space, meaning the tentative trend could change or disappear with further observations; (3) we do not investigate the reliability of the isochrone ages for the \citet{Botelho_2020} solar twin sample, particularly for the middle-aged stars, which again could alter our interpretation of these abundance-age trends. 

\begin{deluxetable*}{lcccccc}
\tablecaption{Stellar Rotation Periods \& Ages (Isochrone vs. Gyrochrone) \label{tab:ages_table}}
\tablehead{
\colhead{} &
\colhead{K2-141} &
\colhead{NGTS-10} &
\colhead{WASP-43} &
\colhead{WASP-69} &
\colhead{WASP-80} &
\colhead{HATS-72}
}
\startdata
 & $^{(1)}$ 15.17 $^{+3.63}_{-2.36}$ & $^{(4)}$ $17.29 \pm 0.008$ & $^{(5)}$ $15.60 \pm 0.4$ & $^{(6)}$ $23.07 \pm 0.16$ & $^{(16)}$ $23.5 \pm 3.0$ & $^{(9)}$ $48.725 \pm 0.015$ \\
$^{(Ref)}$ P$_{rot}$ (days) & $^{(2)}$ $14.03 \pm 0.09$ & --- & $^{(3)}$ $15.25 \pm 0.94$ & $^{(3)}$ $20.87 \pm 1.75$ & $^{(5)}$ $8.1 \pm 0.8$ & --- \\
 & $^{(3)}$ $7.88 \pm 0.29$ & --- & --- & $^{(7)}$ $18.7 \pm 3.5$ & $^{(17)}$ $3.15 \pm 0.01$ & --- \\
\hline
$\bar{P}_{rot}$ (days) & $13.49 \pm 0.15$  & $17.29 \pm 0.008$ & $15.55 \pm 0.64$ & $23.04 \pm 0.28$ & --- & $48.725 \pm 0.015$ \\
\hline
 & $^{(10)}$ $6.30^{+6.60}_{-4.70}$ & $^{(4)}$ $10.4\pm2.5$ & $^{(11)}$ $0.2 \pm 0.6$ & $^{(15)}$ $4.675 \pm 4.012$ & $^{(14)}$ $12.5$ & $^{(9)}$ $12.17^{+0.24}_{-0.45}$ \\
$^{(Ref)}$ Isochrone Age (Gyr) & --- & --- & $^{(12)}$ $2.40 \pm 2.30$ & $^{(13)}$ $11.12 \pm 3.8$ & --- & --- \\
& --- & --- & $^{(13)}$ $11.21 \pm 1.5$ & $^{(14)}$ $0.758$ & --- & --- \\
 & --- & --- & $^{(14)}$ $13.0$ & --- & --- & --- \\
\hline
Isochrone Age (Gyr) & $13.65 _{-0.55} ^{+0.26}$ & $10.07 _{-1.78} ^{+1.75}$ & $13.93 _{-0.13} ^{+0.05}$ &$ 13.97 _{-0.05} ^{+0.02}$ & $13.97 _{-0.06} ^{+0.03}$ & $13.25 _{-0.88} ^{+0.54}$ \\
($^{**}$This Work)  &  &  &  &  &  &  \\
\hline
 & $^{(2)}$ $0.740 \pm 0.360$ & $^{(8)}$ $1.068 \pm 0.025$ & $^{(5)}$ 0.8 & $^{(15)}$ $^{*}1.8 \pm 0.8$ & $^{(5)}$ 0.2 & --- \\
$^{(Ref)}$ Gyrochrone Age (Gyr) & --- & --- & $^{(8)}$ $0.769 \pm 0.028$ & $^{(7)}$ $1.1 \pm 0.15$ & $^{(8)}$ $1.601 \pm 0.202$ & --- \\
\hline
Gyrochrone Age (Gyr) & $0.86 \pm 0.01$ & $1.79 \pm 0.12$ & $0.99 \pm 0.07$ & $2.93 \pm 0.04$ & $0.0691 \pm 0.0003$ & $12.46 \pm 0.01$ \\
($^{**}$This Work)  &  &  &  &  & $0.470 \pm 0.057$ &  \\
\enddata

\tablenotetext{}{References include: (1) \citet{Bonomo_2023}, (2) \citet{Barragan_2018}, (3) \citet{Boyle_2026}, (4) \citet{McCormac_2020}, (5) \citet{Salz_2015}, (6) \citet{Bonomo_2017}, (7) \citet{Anderson_2014}, (8) \citet{Gallet_2020}, (9) \citet{Hartman_2020}, (10) \citet{Malavolta_2018}, (11) \citet{Bonfanti_2015}, (12) \citet{Bonfanti_2016}, (13) \citet{Huang_2022}, (14) \citet{Hamer_2022}, (15) \citet{Allart_2025}, (16) \citet{Penev_2018}, (17) \citet{Triantafillides_2026}. *This is a composite age measurement using numerous techniques including \textit{PARSEC} isochrones, lithium abundances, Ca II H\&K emission, metallicity-age relations, and even rotation-age relationships \citep{Allart_2025}. **Isochrone ages in this work are calculated using the \texttt{isochrone} package \citep{Morton_2015} and the \texttt{emcee} \citep{ForemanMackey_2013} Markov Chain Monte Carlo (MCMC) package. Gyrochronological stellar ages in this work are calculated using the \texttt{GPGyro} code \citep{Lu_2024}. We list two such ages for WASP-80 using the stellar rotation periods from \citet{Triantafillides_2026} and \citet{Salz_2015}, respectively.}
\end{deluxetable*}

\begin{deluxetable*}{ccccccc}\label{tab:stellar_abundances_Kdwarfs}
\tabletypesize{\scriptsize}
\tablewidth{0pt} 
\tablecaption{Fundamental Stellar Parameters and Derived Abundances}
\tablehead{
\colhead{Parameter} & \colhead{K2-141} & \colhead{NGTS-10} & \colhead{WASP-43} & \colhead{WASP-69} & \colhead{WASP-80} & \colhead{HATS-72}}

\startdata
T$_{eff}$ & 4599 $\pm 79^{(3)}$, 4472 $\pm 150^{(9)}$ & 4600$\pm 150^{(4)}$ & 4400$\pm 100^{(5)}$, 4184$\pm 116^{(9)}$  & 4700$\pm 50^{(6)}$, 4876$\pm 77^{(2)}$, 4693$\pm 150^{(9)}$ & 4143$\pm 94^{(7)}$ & 4656$\pm 9^{(8)}$ \\
log g & 4.62 $\pm 0.03^{(3)}$,  4.64 $\pm 0.11^{(9)}$ & 4.59$\pm 0.02^{(4)}$ & 4.50$\pm 0.20^{(5)}$, 4.61$\pm 0.14^{(9)}$ & 4.50$\pm 0.15^{(6)}$, 4.50$\pm 0.09^{(2)}$, 4.46$\pm 0.15^{(9)}$ & 4.66$\pm 0.02^{(7)}$ & 4.59$\pm 0.01^{(8)}$ \\
$[$Fe/H$]$ & -0.06$\pm 0.10^{(3)}$, $+0.07 \pm 0.15^{(9)}$ & -0.02$\pm 0.12^{(4)}$ & $-0.01\pm 0.15^{(5)}$, $+0.29 \pm 0.22^{(9)}$ & +0.15$\pm 0.08^{(6)}$, $+0.40 \pm 0.02^{(2)}$, $+0.38 \pm 0.09^{(9)}$ & -0.13$\pm 0.17^{(7)}$ & +0.10$\pm 0.01^{(8)}$ \\
$[$C/H$]$ & $+0.04 \pm 0.09^{(1)}$ & --- & --- & $ +0.40 \pm 0.06^{(1)}$, $ +0.27 \pm 0.05^{(2)}$ & --- & --- \\
$[$O/H$]$ & $-0.04 \pm 0.09^{(1)}$ & --- & $+0.30 \pm 0.09^{(1)}$ & $+0.14 \pm 0.02^{(1)}$, $+0.18 \pm 0.07^{(2)}$ & --- & --- \\
C/O & $\sim 0.66^{(1)}$ & --- & --- & $\sim 1.00^{(1)}$, $\sim 0.68^{(2)}$ & --- & --- \\
\hline
Adopted (This Work)\\
\hline
T$_{eff}$ & 4571 $\pm 70$ & 4600$\pm 150^{(4)}$ & 4308 $\pm 76$ & 4748$\pm 40$ & 4143$\pm 94^{(7)}$ & 4656$\pm 9^{(8)}$ \\
log g & 4.62 $\pm 0.03$ & 4.59$\pm 0.02^{(4)}$ & 4.55$\pm 0.11$ & 4.50$\pm 0.07$ & 4.66$\pm 0.02^{(7)}$ & 4.59$\pm 0.01^{(8)}$ \\
$[$Fe/H$]$ & -0.03$\pm 0.08$ & -0.02$\pm 0.12^{(4)}$ & $+0.12 \pm 0.12$ & $+0.24 \pm 0.02$ & -0.13$\pm 0.17^{(7)}$ & +0.10$\pm 0.01^{(8)}$ \\
\hline
Derived (This Work)\\
\hline
$[$C/H$]$ ($N_{lines}$) & $-0.03 \pm 0.05$ (19) & $+0.05 \pm 0.06$ (12) & $+0.24 \pm 0.05$ (18) & $+0.26 \pm 0.06$ (10) & $+0.05 \pm 0.06$ (10) & $+0.18 \pm 0.05$ (18) \\ 
$[$O/H$]$ ($N_{lines}$) & $+0.02 \pm 0.08$ (32) & $+0.11 \pm 0.08$ (37) & $+0.11 \pm 0.08$ (46) & $+0.35 \pm 0.09$ (18) & $-0.12 \pm 0.08$ (57) & $+0.20 \pm 0.08$ (29) \\ 
C/O & $0.53 \pm 0.13$ & $0.51 \pm 0.14$ & $0.79 \pm 0.20$ & $0.48 \pm 0.13$ & $0.87 \pm 0.23$ & $0.56 \pm 0.14$ \\
\crat\ ($N_{lines}$) & $51 \pm 12$ (10) & $48 \pm 11$ (3) & $50 \pm 12$ (9) & $38 \pm 10$ (4) & --- & $52 \pm 16$ (3) \\
\enddata

\tablecomments{
This table shows the fundamental stellar parameters and carbon and oxygen abundances for the selected K dwarf sample. The top panel includes literature values for the stellar effective temperature, surface gravity and metallicity [Fe/H]. This list is not exhaustive and compiles values from discovery papers the hosts' companion planets or host star abundance surveys. K dwarfs NGTS-10, WASP-80, and HATS-72 have no published [C/H] or [O/H] abundance ratios according to the Hypatia Catalog \citep{Hinkel_2014}. The middle panel lists adopted fundamental stellar parameters for each target. Where multiple values from different references are shown for a target's effective temperature, surface gravity, or metallicity, we adopt a weighted average of the values. The bottom panel lists the derived abundances and C/O ratios from this work. C/O ratios are calculated from Equations \ref{equ:abundance_ratio} and  \ref{equ:co_uncertainty} using the solar abundances from \citep{Asplund_2021}.\\
\textbf{References:} (1) \citet{da_Silva_2024}; (2) \citet{Polanski_2022}; (3) \citet{Malavolta_2018}; (4) \citet{McCormac_2020}; (5) \citet{Esposito_2017}; (6) \citet{Stassun_2017}; (7) \citet{Triaud_2015}; (8) \citet{Hartman_2020}; (9) \citet{Magrini_2022}. Reference (9) contains the stellar parameters used in reference (1). 
}

\end{deluxetable*}

\section{Conclusion}
This work demonstrates the viability of using near-infrared CO and OH lines to derive elemental carbon and oxygen abundances in K dwarf exoplanet host stars, revealing sub-solar to super-solar C/O ratios across the sample. This chemical abundance analysis has yielded novel [C/H] and [O/H] abundance ratios for K dwarf host stars NGTS-10, WASP-80, and HATS-72 and additional carbon and oxygen constraints for K2-141 and WASP-43.

While initial efforts to extract isotopic ratios (\crat, \orat) from $^{13}$C$^{16}$O and $^{12}$C$^{18}$O lines in the K band showed promise, the weak signal-to-noise and challenging normalization of these features limit their immediate utility and are perhaps best measured from M-band CO isotopologue features. Nevertheless, high-resolution NIR spectroscopy will enable robust elemental and possibly isotopic abundance measurements of cool dwarf stars using new and archival data. 

Expanding this approach to a larger sample of K dwarfs hosting high-priority exoplanets will establish a much-needed baseline of host-star composition for interpreting exoplanetary atmospheric data, particularly from current and upcoming JWST and ground-based campaigns (e.g, ESO SupJup Survey Collaboration; KPIC Collaboration; and other high-resolution, cross-correlation spectroscopy studies of exoplanets).

\subsection{Beyond C/O Ratios}
This work focuses primarily on the elemental and isotopic ratios of carbon and oxygen because carbon- and oxygen-bearing molecules are the most commonly detected in the atmospheres of giant exoplanets. Atmospheric retrievals on JWST NIRCam/MIRI/NIRSpec data from the exoplanet companions to our K dwarf host stars have detected and constrained the volume mixing ratios for different molecules including \water, CO, \coo, \methane, and most notably, nitrogen- and sulfur-bearing molecules like \ammonia, \soo and \css (NGTS-10 b: \citet{Parmentier_2026}; WASP-43 b: \citet{Yang_2024}; WASP-69 b: \citet{Guilluy_2022, Schlawin_2024}; WASP-80 b: \citet{Triantafillides_2026}).

Multiple studies find that complementary stellar and exoplanetary C/O ratios alone provide non-unique formation and accretion scenarios \citep{Reggiani_2022, Coria_2024}, but C/O C/N, N/O, S/N, and \crat\ ratios combined can break degeneracies in planet's formation. migration, and accretion histories \citep{Oberg_2011, Bitsch_2022, Pacetti_2022, Crossfield_2023, Coria_2024, Xuan_2026}. 

Complementary host star nitrogen abundances are particularly lacking for K/M dwarf stars as both atomic nitrogen and molecular NH lines at optical wavelengths are inaccessible due to strong molecular absorption. A recent study \citep{Hejazi_2023} has shown that [N/H] can be derived for late K dwarf stars in the H band using CN lines, but it is yet unclear whether this technique applies to cooler M dwarfs. Stellar sulfur abundance measurements are challenging for the similar reasons; however, they may may be possible using Y-band atomic sulfur lines, H+K band \sulfide lines or L+M band SO lines by addressing the difficult continuum/pseudocontinuum normalization or lack of high-quality near- to mid-infrared (Y, L and M band) spectra for these cool dwarf stars.

Regarding exoplanet abundance constraints, \crat\ ratios remain largely unexplored although retrievals on both space- and ground-based observations have demonstrated their capability of providing constraints \citep{Line_2021, Zhang_2021a, Zhang_2021b, Gonzalez_Picos_2025, Costes_2024, Hood_2024, de_Regt_2024, Gandhi_2023, Finnerty_2024, Gonzalez_Picos_2026}. 

These complementary stellar elemental abundance measurements provide crucial context for interpreting the atmospheric compositions and formation pathways of their giant planet companions and ultimately may provide evidence for outer disk formation and accretion for these hot Jupiters.

\begin{acknowledgments}
The authors thank the anonymous referee for their thorough review and invaluable discussion, which substantially improved this manuscript. The authors would also like to thank Chiaki Kobayashi and Dar\'{i}o Gonz\'{a}lez Picos for providing their GCE models and M dwarf \crat\ measurements, respectively. D. C. would like to acknowledge Yuxi (Lucy) Lu for their attentive help in using and troubleshooting the \texttt{GPGyro} code for determining gyrochronological ages.

D. C. and D. D. would like to thank the Robert E. Young Origins of the Universe Chair fund for its generous support. This work was supported in part by the National Science Foundation (NSF) under grant AST-2441920.

This work has made use of data from the European Space Agency (ESA) mission
{\it Gaia} (\url{https://www.cosmos.esa.int/gaia}), processed by the {\it Gaia}
Data Processing and Analysis Consortium (DPAC,
\url{https://www.cosmos.esa.int/web/gaia/dpac/consortium}). Funding for the DPAC
has been provided by national institutions, in particular the institutions
participating in the {\it Gaia} Multilateral Agreement. 

This research has made use of the Astrophysics Data System, funded by NASA under Cooperative Agreement 80NSSC21M00561. This research has made use of the NASA Exoplanet Archive, which is operated by the California Institute of Technology, under contract with the National Aeronautics and Space Administration under the Exoplanet Exploration Program. N.H. acknowledges support from the Rubin/Chile Joint Research Program, funded by the U.S. National Science Foundation (NSF), and thanks the University of Tarapac{\'a} for hosting this research.

This work used The Immersion Grating Infrared Spectrometer (IGRINS) was developed under a collaboration between the University of Texas at Austin and the Korea Astronomy and Space Science Institute (KASI) with the financial support of the US National Science Foundation under grants AST-1229522, AST-1702267 and AST-1908892, McDonald Observatory of the University of Texas at Austin, the Korean GMT Project of KASI, the Mt. Cuba Astronomical Foundation and Gemini Observatory. The RRISA is maintained by the IGRINS Team with support from McDonald Observatory of the University of Texas at Austin and the US National Science Foundation under grant AST-1908892.
\end{acknowledgments}

\facilities{Gemini:South, LDT}

\appendix
\restartappendixnumbering
\section{OH, CO, and $^{13}$CO Line List: Normalization and Fit Parameters}

\begin{deluxetable*}{lllll}[h!]
\tablecaption{Column Descriptions for Supplementary \texttt{datafileA1.txt} \label{tab:a1}}
\tablewidth{0pt}
\tablehead{
  \colhead{Bytes} &
  \colhead{Format} &
  \colhead{Units} &
  \colhead{Label} &
  \colhead{Explanations}
}
\startdata
1\phn--\phn9   & A9    & --- & Species    & Line species (OH, CO, or $^{13}$CO)                            \\
11--19         & F9.3  & \AA     & Center     & Line center wavelength                                         \\
21             & I1    & --- & Flag       & Number of exclusion zones for continuum selection              \\
23--31         & F9.3  & \AA     & MinLamCont & Minimum wavelength of the continuum selection window           \\
33--41         & F9.3  & \AA     & MaxLamCont & Maximum wavelength of the continuum selection window           \\
43--52         & F10.3 & \AA     & MinLamExc1 & Minimum wavelength of exclusion window 1                       \\
54--63         & F10.3 & \AA     & MaxLamExc1 & Maximum wavelength of exclusion window 1                       \\
65--74         & F10.3 & \AA     & MinLamExc2 & Minimum wavelength of exclusion window 2; '999999' if Flag == '1'                       \\
76--85         & F10.3 & \AA     & MaxLamExc2 & Maximum wavelength of exclusion window 2; '999999' if Flag == '1'                       \\
87--95         & F9.3  & \AA     & MinLamFit  & Minimum wavelength of the $\chi^2$ abundance fit window        \\
97--105        & F9.3  & \AA     & MaxLamFit  & Maximum wavelength of the $\chi^2$ abundance fit window        \\
\enddata
\tablecomments{Wavelength units are \AA\ ($= 0.1$~nm). ---\ denotes a dimensionless quantity.
Flag values indicate the number of exclusion zones applied by the continuum selection
routine (0, 1, or 2). The cell value "999999" is a null value.}
\end{deluxetable*}

\section{Line-by-Line Relative Abundances}

\begin{figure*}[h!]
    \includegraphics[width=1.0\textwidth]{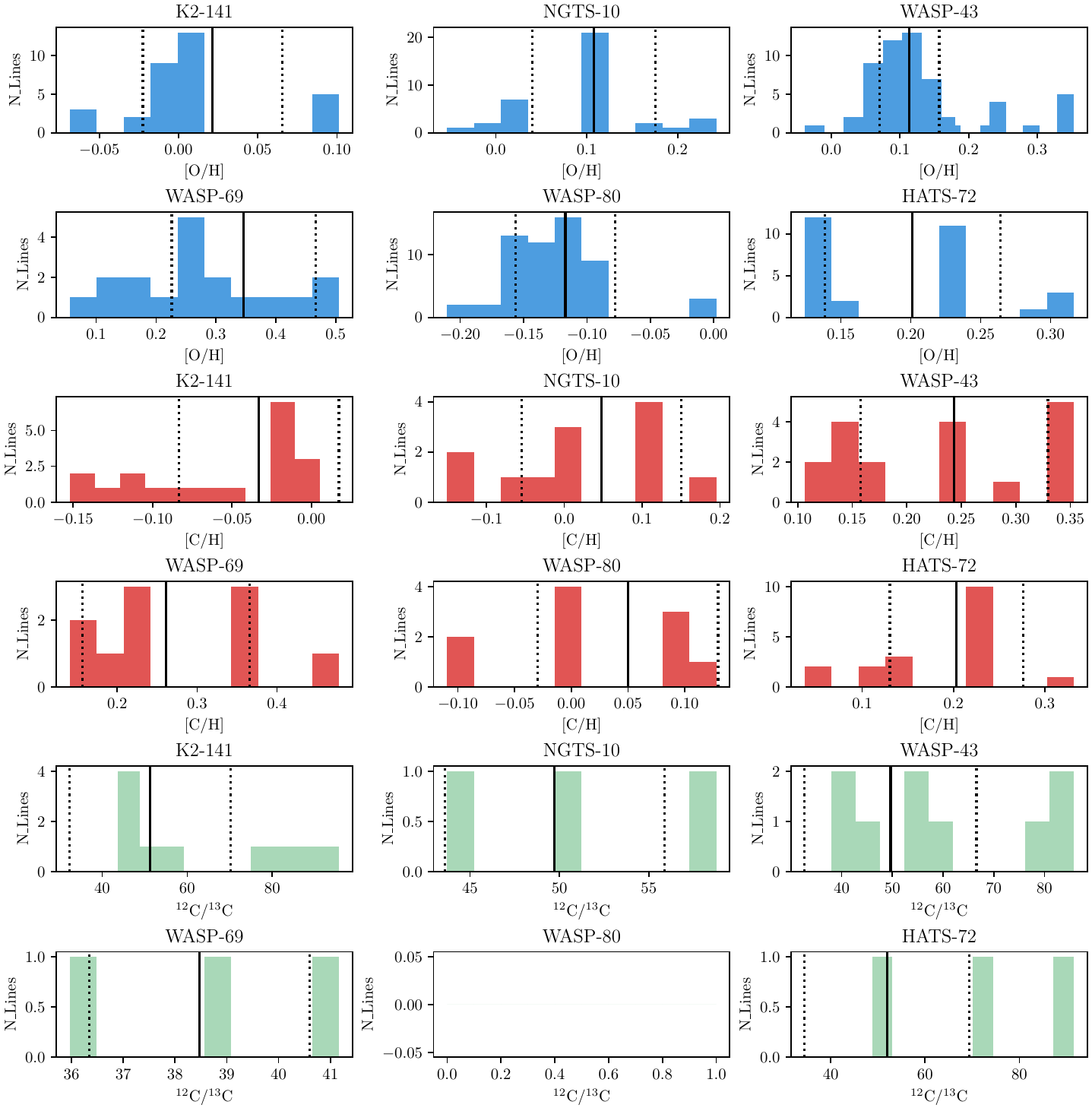}
    
    \caption{Histograms for line-by-line abundance measurements for oxygen (top, blue), carbon (middle, red), and \crat\ (bottom, green) The solid black lines represent the weighted average of the abundance values, weighted by statistical uncertainty ($\Delta \chi^2 = 1$) for each measurement. The dotted lines represent the standard deviation ($\pm 1 \sigma$) bounds. We are unable to determine the \crat\ ratio for WASP-80. Final [O/H], [C/H], and \crat\ ratios are given in Table \ref{tab:stellar_abundances_Kdwarfs}.
    }
     \label{fig:abundance_hist}
\end{figure*}



\bibliography{kdwarfs}{}
\bibliographystyle{aasjournalv7}



\end{document}